\documentclass[aip, amsmath,amssymb,reprint,graphicx]{revtex4-1}
\usepackage{graphicx,color}
\usepackage{dcolumn}
\usepackage{bm}
\usepackage{graphicx,titlesec,amsthm}
\usepackage{extarrows}
\usepackage[colorlinks,urlcolor=blue,citecolor=blue]{hyperref}

\newcommand{\ii}{\mathrm{i}}
\newcommand{\sn}{\mathrm{sn}}
\newcommand{\cn}{\mathrm{cn}}
\newcommand{\dn}{\mathrm{dn}}

\newcommand{\ee}{\mathrm{e}}
\newcommand{\dd}{\mathrm{d}}
\newcommand{\diag}{\mathrm{diag}}

\newcommand{\sech}{\mathrm{sech}}

\newtheorem{definition}{Definition}

\draft 

\usepackage[utf8]{inputenc}
\usepackage[T1]{fontenc}
\usepackage{mathptmx}
\usepackage{etoolbox}

\makeatletter
\def\@email#1#2{%
 \endgroup
 \patchcmd{\titleblock@produce}
  {\frontmatter@RRAPformat}
  {\frontmatter@RRAPformat{\produce@RRAP{*#1\href{mailto:#2}{#2}}}\frontmatter@RRAPformat}
  {}{}
}%
\makeatother

\begin{document}

\preprint{AIP/123-QED}
\title{Two types of elliptic dark soliton solutions for the Hirota equation} 

\author{Qiaofeng Huang*}
 
\author{Xuan Sun**}%
 \email{sunx@dhu.edu.cn}
\affiliation{
*School of Mathematics, South China University of Technology, Guangzhou, China, 510641
\\
**School of Mathematics and Statistics, Donghua University, Shanghai, China, 201620}

\date{\today}

\begin{abstract}
We primarily study concave-downward and convex-upward types of elliptic dark soliton solutions for the Hirota equation, exhibiting a concave-downward shape on both upper and lower envelope surfaces and showing a convex-upward shape on the lower envelope surface, respectively. 
By analyzing the supremum and infimum of solutions, we provide the existence conditions for these two types of elliptic dark solitons. Additionally, we study two-elliptic dark soliton solutions combining both types with the same velocity and investigate the elastic collisions between these two types of solutions with different velocities.
\end{abstract}

%\pacs{}% insert suggested PACS numbers in braces on next line

\maketitle %\maketitle must follow title, authors, abstract and \pacs

%\begin{quotation}
%Recently, the nonlinear propagation of elliptic functions has been observed in hydrodynamical and optical experiments, which gives rise to interest in the application of elliptic dark solitons.
%In this work, we present a detailed study of two distinct types of elliptic dark soliton solutions for the Hirota equation, namely the concave-downward type (CD-type) and convex-upward type (CU-type) elliptic dark solitons.  
%We systematically analyze the dynamical behavior of these two types of elliptic dark soliton solutions and establish their corresponding existence conditions.
%Additionally, we demonstrate the two-elliptic dark soliton solutions with the same velocity that maintain stable separation distances during propagation. For different velocities, the collisions between CD-type and CU-type elliptic dark solitons are revealed to be elastic.
%\end{quotation}

\section{Introduction}\label{sec:level1}
Dark solitons, characterized by localized regions of reduced intensity within a background of higher intensity \cite{MR985322, RevModPhys.61.763}, have been investigated for a profound understanding of various complex nonlinear phenomena \cite{PhysRevLett.110.124101,PhysRevLett.83.5198,PhysRevLett.96.245001,kodama1987nonlinear,hosseini2018new,Demontis201561,RevModPhys.68.423,RevModPhys.71.463,MR702914,1967461133} occurring in fluid dynamics, plasma physics, nonlinear optics, Bose-Einstein condensates and so on.
Up to now, there have been a large number of studies on dark soliton solutions of the defocusing nonlinear evolution equations by utilizing Darboux-B\"acklund transformation, Hirota bilinear methods and so on \cite{MR2854923,MR3403397,MR1212008,MR3709441,MR3623400,Lou_2014,MR3663709}.
As widely understood, the Hirota equation is an integrable equation derived by Hirota \cite{hirota1973exact}, which can be regarded as a modified nonlinear Sch\"{o}rdinger equation while the third order dispersion and time-delay changes are taken into account, showing a better property in depicting ultrashort pulses in optical fibers and playing a vital role in the fields of physics \cite{tasgal1992soliton,ankiewicz2010rogue,tao2012multisolitons}. 
Due to the effect of high-order dispersion, Hirota equation possesses more abundant dynamic phenomena \cite{Zhang_2021,PhysRevE.104.014201,MR4670721}, such as the double valley dark soliton. Nowadays, the research on the plane wave background has become relatively mature. 

Significant attention has been devoted to solutions of integrable nonlinear equations on elliptic function backgrounds \cite{LiG-20-SG,MR4844570,MR4768161,hoefer2023kdv,zhen2023rogue,chen2024bright,rao2024dark,chen2018rogue1,chen2018rogue2,feng2020multi,ling2023elliptic,ling2023multi,mucalica2024dark,8189835,MR2132713}, which have been observed in photonic crystal fibers and nonlinear metamaterials \cite{Russell-2003,LapineSK-2014}.
However, due to the difficulty of elliptic functions, related studies on solutions for the Hirota equation on elliptic function backgrounds are still limited.
Rogue waves of the focusing Hirota equation on elliptic function backgrounds have been presented \cite{peng2020characteristics,gao2020rogue}.
 A recent study presents elliptic dark soliton solutions of the defocusing Hirota equation \cite{wang2024dark}, which exhibits the single elliptic dark solitons with the concave downward shape on both upper and lower envelope surfaces. 
 We call these solutions the concave-downward type (CD-type) elliptic dark soliton solutions.
Naturally, we wonder if there exist elliptic dark soliton solutions with different structures in the Hirota equation.
It motivates us to systematically explore the dynamic behaviors of elliptic soliton solutions for the Hirota equation and demonstrate these interesting phenomena. 

In this paper, we present the explicit single elliptic dark soliton solutions of the Hirota equation, which exhibit two different types of solitons.
Besides the CD-type elliptic dark soliton solutions, we provide elliptic dark solitons exhibiting concave downward on the upper envelope surface and convex upward on the lower envelope surface, called the convex-upward type (CU-type) elliptic dark soliton solutions.
Furthermore, we provide the corresponding existence conditions of these two types of elliptic dark soliton solutions. 
As far as we know, this has not been reported before. Meanwhile, the conditions that allow two types of the single dark soliton solutions to move at the same velocity are provided.
Furthermore, we explore the collision dynamics of multi-elliptic dark solitons, which reveals that the interaction between two types of the single dark soliton solutions results in an elastic collision.

The structure of this paper is outlined as follows: In section \ref{sec2}, we derive the exact single elliptic dark solutions in terms of theta functions for the Hirota equation. 
Moreover, we demonstrate the CD-type and CU-type elliptic dark soliton solutions by presenting their corresponding supremum and infimum.
The relationship between the spectral parameter and the dynamical behaviors of the single elliptic dark solutions is also presented.
In section \ref{sec3}, based on the $N$-fold Darboux–B\"{a}cklund transformation, we derive the multi-elliptic dark soliton solutions of the Hirota equation. Furthermore, we reveal the elastic collisions among these dark solitons by analyzing the asymptotic expressions of solutions along the trajectories of dark solitons and in the region between two dark solitons. The conclusions are given in section \ref{conclusion}.

\section{Single elliptic dark solutions}\label{sec2}
In this work, we consider the Hirota equation
\begin{equation}\label{hirota}
\ii \psi_t+\epsilon_1(\psi_{xx}-|\psi|^2\psi)+\ii \epsilon_2\left(\psi_{xxx}-3|\psi|^2 \psi_x\right)=0,
\end{equation}
where $(x,t)\in \mathbb{R}^2$ are spatial distribution and time evolution coordinates, $\psi=\psi(x,t)\in \mathbb{C}$ is the envelope of the wave field, and
parameters $\epsilon_{1,2}$ are two real numbers.
When $\epsilon_1=0$ and $\epsilon_2\neq 0$, the Hirota equation \eqref{hirota} would deduce into the modified Korteweg-de Vries equation;
when we let $\epsilon_1\neq 0$ and $\epsilon_2=0$, the Hirota equation \eqref{hirota} would convert into the nonlinear Sch\"{o}rdinger equation.
The Lax pair of the Hirota equation \eqref{hirota} is expressed as
\begin{equation}\label{lax}
   \!\! \Phi_x=\mathbf{U}(\lambda;\mathbf{Q})\Phi,\quad
    \Phi_t=\mathbf{V}(\lambda;\mathbf{Q})\Phi,\quad
    \mathbf{Q}=\begin{bmatrix} 0 & \psi \\ \psi^* &0 \end{bmatrix}\!,\!\!
\end{equation}
where $\Phi=\Phi(x,t;\lambda)$, $\mathbf{U}=\mathbf{U}(\lambda;\mathbf{Q})=\sqrt{2}\left(\ii\lambda \sigma_3+\mathbf{Q}\right)/2$, $\mathbf{V}=\mathbf{V}(\lambda;\mathbf{Q})=\epsilon_2(2\lambda^2\mathbf{U}-\ii\lambda\sigma_3\mathbf{Q}_x+\mathbf{Q}_x\mathbf{Q}/2-\mathbf{Q}\mathbf{Q}_x/2) +\sqrt{2}\epsilon_2(\ii\lambda\sigma_3\mathbf{Q}^2+\mathbf{Q}^3-\mathbf{Q}_{xx})/2
    -\epsilon_1(\ii \sigma_3\mathbf{Q}^2/2+\sqrt{2}\lambda\mathbf{U}-\sqrt{2}\ii \sigma_3\mathbf{Q}_x/2)$, $\sigma_3:=\diag(1,-1)$ is called the third Pauli matrix and the region of spectral parameter $\lambda$ is $(\mathbb{C}\cup\{\infty\})$.
Moreover, the compatibility condition of the linear system \eqref{lax}: $\Phi_{xt}=\Phi_{tx}$ is equivalent to the Hirota equation \eqref{hirota}. 
Thus, in the following contexts, we are focusing on the properties of the Lax pair, which is useful in studying solutions of the Hirota equation \eqref{hirota}.

\subsection{Construction of solutions}
The exact solutions play a crucial role in researching and revealing different types of elliptic dark soliton solutions.
Define the matrix $\mathbf{L}\equiv\mathbf{L}(x,t;\lambda)$ satisfying the stationary zero curvature equations
\begin{equation}\label{zero}
    \mathbf{L}_x=[\mathbf{U},\mathbf{L}],\quad
    \mathbf{L}_t=[\mathbf{V},\mathbf{L}],
\end{equation}
with commutator defined by $[\mathbf{A}, \mathbf{B}]=\mathbf{A}\mathbf{B}-\mathbf{B}\mathbf{A}$, the compatibility condition $\mathbf{L}_{xt}= \mathbf{L}_{tx}$ of which also gives the Hirota equation \eqref{hirota}. Building upon \cite{ling2023stability}, we assume that the matrix function $\mathbf{L}$ is a quadratic polynomial of $\lambda$:
$\mathbf{L}=\mathbf{L}_0(x,t)\lambda^2+\mathbf{L}_1(x,t)\lambda+\mathbf{L}_2(x,t)$.
Plugging it into Eq. \eqref{zero} and comparing the coefficient of $\lambda$, we could obtain
\begin{equation}\label{g}
    \mathbf{L}
    =\begin{bmatrix}
        \lambda^2+|\psi|^2/2+b_3 & \ii \psi(\lambda-\mu)\\
        -\ii\psi^*(\lambda-\mu^*) & -(\lambda^2+|\psi|^2/2+b_3)
    \end{bmatrix} ,
\end{equation}
where $\mu=\sqrt{2}\ii (\ln \psi)_x/2$ and
$b_3\in \mathbb{R}$ satisfies the equation
\begin{equation}\label{f1}
    \mathbf{Q}_t+2\epsilon_2 b_3\mathbf{Q}_x-2\ii \epsilon_1b_3\mathbf{Q}\sigma_3-\epsilon_2[\mathbf{Q}_x\mathbf{Q},\mathbf{Q}]=0.
\end{equation}
Then, the determinant of the matrix $\mathbf{L}$ is
\begin{equation}\label{f10}
    \det(\mathbf{L})=-L_{11}^2-L_{12}L_{21}=-\sum_{i=0}^{4}\lambda^is_{4-i}\equiv P(\lambda),
\end{equation}
where $s_0=1$, $s_{1,2,3,4}$ are real constants and satisfying $b_3=s_2/2$, $\mu+\mu^*=s_3/v$, $\mu\mu^*=[(s_2/2+v/2)^2-s_4]/v$, $v=|\psi|^2$.
Functions $\mu$, $\mu^*$ can be solved exactly:
\begin{equation}\label{f2}
\begin{split}
   & \mu,\mu^*=(s_3\pm \ii \sqrt{R(v)})/(2v),\\
   &R(v)=v^3+2s_2v^2+(s_2^2-4s_4)v-s_3^2.
\end{split}
\end{equation}
Consider the traveling wave transformation
    \begin{equation}\label{eq:transformation}
        (x,t)\xlongequal[\eta=t]{\xi=x- \epsilon_2 s_2 t}  (\xi,\eta).
    \end{equation}
In the following, we would consider solutions under the $(\xi,\eta)$-coordinate system.
Combined with the above analysis, the modulus square of solutions for the Hirota equation \eqref{hirota} could be represented as
    \begin{equation}\label{f9}
        |\psi|^2=v=2k^2\alpha^2\left(\sn^2(\alpha\xi)-\sn^2(4\ii l K)\right),
    \end{equation}
    where $\alpha=\sqrt{(v_3-v_1)/2}$, $l\in[0,-\ii\tau/4)$, $\tau=\ii K^{\prime} /K$, $k=\sqrt{(v_2-v_1)/(v_3-v_1)}$ is called the modulus, $K\equiv K(k)$, $K^{\prime}\equiv K(\sqrt{1-k^2})$ are the complete elliptic integrals,  and $v_{1,2,3}$ are parameterized by $v_1=-2\alpha^2 k^2 \sn^2(4\ii l K)$, $v_2=2\alpha^2 k^2 \cn^2(4\ii l K)$, $v_3=2\alpha^2 \dn^2(4\ii l K)$. 
In combination with Eqs. \eqref{f1} and \eqref{f2}, we could obtain the solutions for the Hirota equation \eqref{hirota} by integration. According to the relationship between theta functions and Jacobi elliptic functions (\cite{gradshteyn2014table} p.888), the exact expression of solutions in terms of theta functions for the Hirota equation \eqref{hirota} can be presented as follows:
\begin{equation}\label{solu2}
         \psi=\chi\frac{\vartheta_1(2\ii l+\hat{\alpha}\xi)}{\vartheta_4(\hat{\alpha}\xi)}\ee^{\omega_1\xi+\omega_2\eta},\quad \chi=\frac{\sqrt{2}\ii\alpha\vartheta_2\vartheta_4}{\vartheta_3\vartheta_4(2\ii l)},
     \end{equation}
     where the transformation between $(\xi,\eta)$ and $(x,t)$ is defined in Eq. \eqref{eq:transformation}, $\hat{\alpha}=\alpha/(2K)$, $ \omega_1=-\alpha Z(4\ii l K)$, $\omega_2=\ii(\epsilon_1 s_2-\sqrt{2} \epsilon_2 s_3)$, $Z(4\ii l K)\in \ii \mathbb{R}$ is called the Jacobi Zeta function. The proof details of Eq. \eqref{solu2} are presented in Appendix. \ref{app1}. In particular, when $l=0$, the solution $\psi$ of the Hirota equation \eqref{hirota} reduces into a sn-type solution: $\psi=\sqrt{2}\ii\alpha k\sn(\alpha\xi)\ee^{\ii \epsilon_1 s_2\eta}$. As $k\rightarrow 1^{-}$, it degenerates into a general dark soliton solution.
 
We proceed to obtain the solution of the Lax pair \eqref{lax} under the coordinate transformation \eqref{eq:transformation}. 
Without loss of generality, we set $\pm y$ as two eigenvalues of the matrix $\mathbf{L}$ under the $(\xi,\eta)$ coordinate, which means $-y^2=\det{(\mathbf{L})}$.
It is easy to verify that $(1,r_1)^{\top}$ and $(1,r_2)^{\top}$ are the kernels of the matrices $y \mathbb{I}-\mathbf{L}$ and $-y \mathbb{I}-\mathbf{L}$ respectively, where $\mathbb{I}$ is the $2\times 2$ identity matrix and $r_i=r_i(\xi,\eta;\lambda)=((-1)^{i+1}y-L_{11})/L_{12}=L_{21}/((-1)^{i+1}y+L_{11})$ with $i=1,2$. 
 In addition, vectors $\hat{\Phi}(1,r_1)^{\top}$ and $\hat{\Phi}(1,r_2)^{\top}$ are also kernels of the matrices $y \mathbb{I}-\mathbf{L}$ and $-y \mathbb{I}-\mathbf{L}$ respectively, where $\hat{\Phi}=\hat{\Phi}(\xi,\eta;\lambda)$ is a fundamental matrix solution of the Lax pair under the coordinate transformation \eqref{eq:transformation} with the initial data $\hat{\Phi}(0,0;\lambda)=\mathbb{I}$. 
 Based on this, we obtain $\Phi_{2i}=r_i\Phi_{1i}$, $i=1,2$, one could refer to \cite{feng2020multi} for the detailed calculation, where $\Phi_{ij}$ represents the $(i,j)$-element of $\Phi$. Subsequently, we could derive the fundamental solution $\Phi$ of the Lax pair for the Hirota equation \eqref{hirota} by integration. In the following, we intend to represent the solution $\Phi$ by using theta functions. From Eqs. \eqref{zero}, \eqref{g} and setting $\lambda=\mu$, we obtain the derivative of $L_{12}$ with respect to $\xi$ in two different ways: $L_{12,\xi}=\ii \psi \mu_{\xi},\quad L_{12,\xi}=-\sqrt{2}\psi L_{11}$, which implies $\mu_{\xi}^2=-2L_{11}^2=2P(\mu)$. 
 Functions $\lambda$ and $y$ can be parameterized by the uniform parameter $z$ as follows
\begin{equation}\label{f23}
    \lambda(z)=\mu\left(\frac{z_l}{\alpha}\right),\quad y(z)=\frac{\sqrt{2}\alpha}{4K}\frac{\dd}{\dd z}\mu\left(\frac{z_l}{\alpha}\right),
\end{equation}
where $z_l=2\ii (z-l)K$.
Combining with the Eqs. \eqref{f2} and \eqref{f9}, we can express $\lambda(z)$ as
\begin{equation}\label{lambda}
    \lambda(z)=\frac{ \sqrt{2}\ii \alpha}{2}\frac{-\mathrm{scd}(4\ii l K)+\mathrm{scd}(z_l)}{\sn^2(z_l)-\sn^2(4\ii l K)},
\end{equation}
where $\mathrm{scd}(\cdot)=\sn(\cdot)\cn(\cdot)\dn(\cdot)$. 
Specifically, by applying the similar method illustrated in \cite{LING2024108866}, we know that the function $\lambda(z)$ is a conformal mapping that maps the rectangular region
\begin{equation}\label{f16}
    S:=\left\{z\in \mathbb{C}| | \Re(z)-l|\leq -\ii\tau/2,\,\, 0\leq \Im(z)\leq 1/2 \right\}
\end{equation}
onto the entire complex plane $\mathbb{C}\cup \{\infty\}$ except two cuts. 

Based on the above analysis, the fundamental solution of the Lax pair \eqref{lax} with respect to the solution $\psi$ given by Eq. \eqref{solu2} can be expressed in terms of theta functions:
 \begin{equation}\label{psi}
        \Phi=\chi\frac{\vartheta_4(2\ii l )}{\vartheta_4\left(\hat{\alpha}\xi \right)}\Lambda
        \begin{bmatrix}
             \frac{ \vartheta_1\left(\hat{\alpha}\xi -\ii(z-l) \right)}{\vartheta_4(-\ii(z-l) )} &
             \frac{ \vartheta_4\left(\hat{\alpha}\xi +\ii(z+l) \right)}{\vartheta_1(\ii(z+l) )}\\
              \frac{ \vartheta_4\left(\hat{\alpha}\xi -\ii(z+l) \right)}{\vartheta_1(-\ii(z+l) )}&
              \frac{ \vartheta_1\left(\hat{\alpha}\xi +\ii(z-l) \right)}{\vartheta_4(\ii(z-l) )}     
        \end{bmatrix}\mathbf{E},
    \end{equation}
    where 
    the transformation between $(\xi,\eta)$ and $(x,t)$ is defined in Eq. \eqref{eq:transformation},
    $\Lambda=\mathrm{diag}(1,\ii \ee^{-\omega_1\xi-\omega_2\eta})$, $\omega_1$ and $\omega_2$ are defined in Eq. \eqref{solu2}; $\mathbf{E}=\diag(E_1(\xi,\eta;z),E_2(\xi,\eta;z))$ and $E_{i}(\xi,\eta;z)=\ee^{W_i(z)\xi+V_i(z)\eta}$
    with $W_1(z)=-\sqrt{2}\ii\lambda/2+\alpha Z(z_l)$, $W_2(z)=-\sqrt{2}\ii\lambda/2+\ii\hat{\alpha} +\alpha Z(\ii K^{\prime}-4\ii lK-z_l)$, and
      $V_{1,2}(z)=\ii(\epsilon_1s_2-\sqrt{2}\epsilon_2s_3)/2 \pm \ii(\sqrt{2}\epsilon_2\lambda-\epsilon_1)y$. For the elaborate proof of Eq. \eqref{psi}, one could turn to Appendix \ref{app1}.

Subsequently, we intend to use the fundamental solution \eqref{psi} to construct single elliptic dark soliton solutions of the Hirota equation \eqref{hirota} through developing the generalized Darboux-B\"{a}cklund transformation as presented in \cite{feng2020multi}.  The Darboux matrix can be represented as
\begin{align}
    \mathbf{T}_1=\mathbb{I}-\frac{\lambda_1-\lambda_1^{*}}{\lambda-\lambda_1^{*}}\frac{\Phi_1\Phi_1^{\dagger}\sigma_3}{\Phi_1^{\dagger}\sigma_3\Phi_1},
\end{align}
where $\Phi_1=\Phi(\xi,\eta;\lambda_1)\mathbf{c}_1$, $\Phi(\xi,\eta;\lambda_1)=[\phi_1,\phi_2]$ is given by Eq. \eqref{psi} with $\lambda=\lambda_1$, $\mathbf{c}_1=[c_{11},c_{12}]^{\top}$ and $c_{11,12}\in \mathbb{C}$.
Together with the Lax pair in Eq. \eqref{lax}, we know that the corresponding spectral problem is self-adjoint, which restricts the spectral parameter must satisfy $\lambda\in \mathbb{R}$. It can be noticed that the matrix $\mathbf{T}_1$ degenerates into the identity matrix $\mathbb{I}$ since $\lambda_1\in \mathbb{R}$. Under this case, the matrix $\mathbf{T}_1$ is no longer dependent on $\lambda$, which implies that we could not utilize it to generate other different solutions. 

Therefore, we introduce some special parameters $\mathbf{c}_1$ such that the parameter $c_{12}$ is dependent on $\lambda_1-\lambda_1^{*}$ to ensure the matrix $\mathbf{T}_1$ relating to the spectral parameter $\lambda$. If $c_{11}=1$ and $c_{12}=a_1(\lambda_1-\lambda_1^{*})$ with $a_1\in\mathbb{C}$, the Darboux matrix $\mathbf{T}_1(\xi,\eta;\lambda)\neq \mathbb{I}$. Utilizing such $\mathbf{T}_1$, we could construct the single elliptic dark soliton solutions.  

When the parameter $z_1\in S$ satisfies $\Re(z_1)=\pm\ii\tau/4$, the explicit expression of the single elliptic dark soliton solution for the Hirota equation \eqref{hirota} is
\begin{equation}\label{psi1}
 \!\! \psi^{[1]}=\chi\frac{\vartheta_1(\hat{\alpha}\xi +2\ii l  )}{\vartheta_4(\hat{\alpha}\xi )}\cdot\frac{p^{*}_1p^{-1}_1H\hat{E}_1+1}{G\hat{E}_1+1}\ee^{\omega_1\xi+\omega_2\eta}\!,
 \!\!\end{equation}
where the transformation between $(\xi,\eta)$ and $(x,t)$ is defined in Eq. \eqref{eq:transformation}, $G=\vartheta_4(\hat{\alpha}\xi +2z_c )/(2\hat{a}_1\vartheta_4(\hat{\alpha}\xi )\vartheta_1(2z_c ))$, $H=- \vartheta_1(\hat{\alpha}\xi +2z_c +2\ii l  )/( 2\hat{a}_1\vartheta_1(\hat{\alpha}\xi +2\ii l  )\vartheta_1(2z_c ))$, $\hat{E}_1=\hat{E}_1(z_1)=\exp(\alpha(-Z(4z_cK)-k^2\sn(\pm\ii K^{\prime}/2+2\ii l K-2z_cK)\sn(\pm\ii K^{\prime}/2+2\ii l K+2z_cK)\sn(4z_cK))\xi+2\ii(\sqrt{2}\lambda-1)y\eta)$, $z_c=\Im(z_1)$, $p_1=\vartheta_4(\ii(z_1-l))/\vartheta_1(-\ii(z_1+l))$ and $\hat{a}_1=\Im(a_1)\ee^{-\ii\tau /4}>0$. The detailed proof of Eq. \eqref{psi1} is given in Appendix. \ref{app3}. As $\xi\rightarrow \infty$, the asymptotic expression of $\psi^{[1]}$ coincides with the exact expression of $\psi$ in Eq. \eqref{solu2}. When $\xi\rightarrow -\infty$, $\psi^{[1]}\rightarrow -\chi p_1^{*}p_1^{-1}\vartheta_1(\hat{\alpha}\xi+2z_c+2\ii l)/\vartheta_4(\hat{\alpha}\xi+2z_c)\ee^{\omega_1\xi+\omega_2\eta}$. It can be seen that the asymptotics on both sides of the single soliton differ by a phase shift of $2z_c/\hat{\alpha}$.

 Given the parameters $k$, $l$, $\alpha$, $a_1$ and $z_1$, the solution exhibits two types of single elliptic dark solutions by plugging parameters into Eq. \eqref{psi1}, as shown in Fig. \ref{fig5}(a) and \ref{fig5}(c). 
By taking $\alpha=1$, $k=1/2$, $l=-\ii\tau/20$, $z_1=2\ii/5+\ii\tau/4$, a CD-type elliptic dark soliton solution is shown in Fig. \ref{fig5}(a), which exhibits a concave downward shape on both the upper and lower envelope surfaces. 
Changing the parameter $z_1=-\ii/6-\ii\tau/4$, we obtain a CU-type elliptic dark soliton solution shown in Fig. \ref{fig5}(c), which exhibits a concave downward shape on the upper envelope surface and a convex upward shape on the lower envelope surface. 
It could be noted that the above two types of elliptic dark soliton solutions exhibit distinct structures on the lower envelope surface with different parameters.
In the following, we are going to study these two types of elliptic dark soliton solutions and explore their existence conditions.

\subsection{Two types of single elliptic dark soliton solutions}
In this subsection, we focus on the above two different dynamic behaviors of the single elliptic dark solitons and illustrate the correspondence between parameters we choose and dynamic behaviors they exhibit. As analyzed in the previous subsection, the CD-type and the CU-type of single elliptic dark soliton solutions exhibit different dynamical behaviors on the lower envelope surface, namely concave downward and convex upward respectively.
  
In order to study these dynamical behaviors, it is natural for us to explore the supremum and infimum of $\psi^{[1]}$ through the specific expression given in Eq. \eqref{psi1}. When we calculate its infimum, it can be achieved by taking the maximum value for the denominator and the minimum value for the numerator. For the parameters $l$ and $z_1$ with different real parts, the infimum of $\psi^{[1]}$ can be expressed in different forms. Through the detailed proof given in Appendix \ref{app3}, we can obtain the following results.
For $l\neq 0$, we can get the following infimum of $\psi^{[1]}$:
    \begin{itemize}
        \item When $z_1\in S$ with $\Re(z_1)=\ii\tau/4$:
   \begin{equation}\label{inf1}
       \begin{split}
     \!\!\!   \psi_{\mathrm{inf}}^{[1]}\!=&\chi\frac{\vartheta_1(2\ii l )(\!B_1\!+\!1\!+\!(\!B_1\!-\!1\!)\!\tanh(\!\hat{L}\!-\!\rho_1\!)\!)}{2\vartheta_3}\ee^{\omega_1\xi+\omega_2\eta},
        \end{split}
        \end{equation}
        which demonstrates the dynamical behavior of a dark soliton.
        \item When $z_1\in S$ with $\Re(z_1)=-\ii\tau/4$:
    \begin{equation}\label{inf2}
        \begin{split}
         \psi_{\mathrm{inf}}^{[1]}=\chi\frac{\vartheta_1(2\ii l )}{\vartheta_3}(1+B_2\mathrm{sech}(\hat{L}-\rho_1))\ee^{\omega_1\xi+\omega_2\eta},
         \end{split}
           \end{equation}
        which demonstrates the dynamical behavior of a bright soliton.
    \end{itemize}
In the above equations, $\rho_1=(\ln2\hat{a}_1\vartheta_1(2 z_c  ))/2$  and $\hat{L}=(\ln\!\hat{E}_1)/2$; $B_1=(d_{m}+\ii\sqrt{|\vartheta_1(2\ii l )|^4-d_m^2})/|\vartheta_1(2\ii l )|^2$, $B_2=\ii\sqrt{(\tilde{d}_{m}-|\vartheta_1(2\ii l )|^2)/\left(2|\vartheta_1(2\ii l )|^2\right)}$ with $d_m=\underset{{\xi\in\mathbb{R}}}{\mathrm{min}}\left(\Re\left(d(\xi)\right)\right)$, $\tilde{d}_{m}=\mathrm{max}[d_m,|\vartheta_1(2\ii l )|^2]$ and $d(\xi)=-p^{*}_1p_1^{-1}\vartheta_1(\hat{\alpha}\xi +2z_c +2\ii l)\vartheta_1(\hat{\alpha}\xi -2\ii l  )$.

Similarly, we can obtain the expression for the supremum.
When $l\neq 0$, $z_1\in S$ with $\Re(z_1)=\pm\ii\tau/4$, the supremum of $\psi^{[1]}$ is
   \begin{equation}\label{sup}
    \begin{split}
     \!\!\!\!\!  \psi^{[1]}_{\sup}\!=\!\chi\frac{\vartheta_2(2\ii l )(\!B_3\!+\!1\!+\!(\!B_3\!-\!1\!)\!\tanh(\!\hat{L}\!-\!\rho_1\!)\!)}{2\vartheta_4}\ee^{\omega_1\xi+\omega_2\eta},
         \end{split}
    \end{equation}
    where $B_3=(d_{M}+\ii\sqrt{|\vartheta_2(2\ii l )|^4-d_M^2})/|\vartheta_2(2\ii l )|^2$ and $d_M=\underset{{\xi\in\mathbb{R}}}{\mathrm{max}}\left(\Re\left(d(\xi)\right)\right)$.

\begin{figure}[htp]
	\centering
	\includegraphics[width=0.4\linewidth]{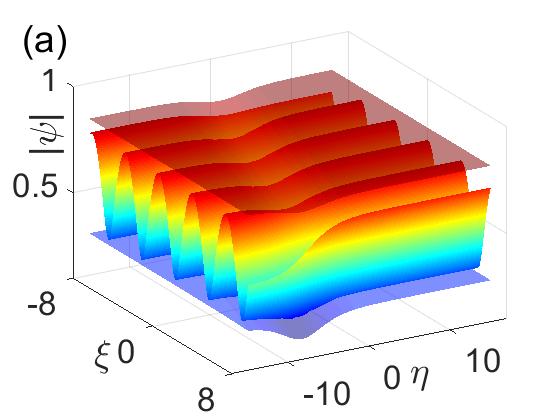}
	\includegraphics[width=0.4\linewidth]{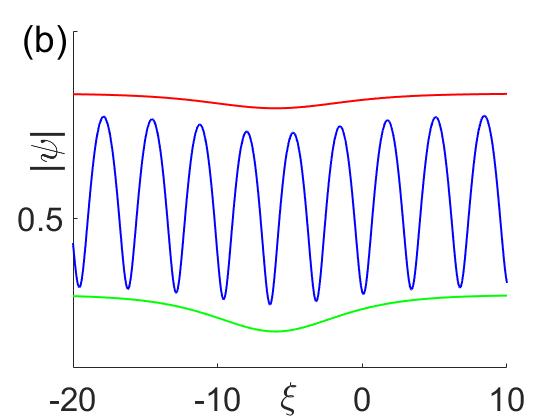}\\
	\includegraphics[width=0.4\linewidth]{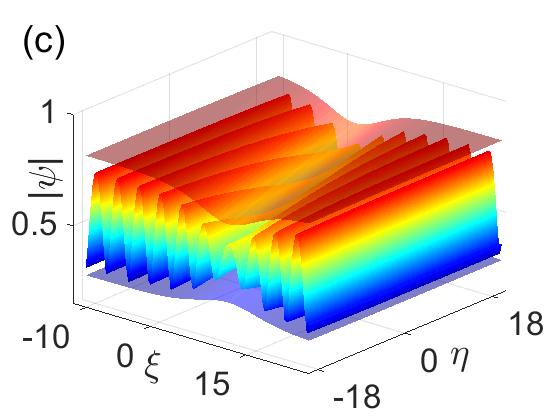}
        \includegraphics[width=0.4\linewidth]{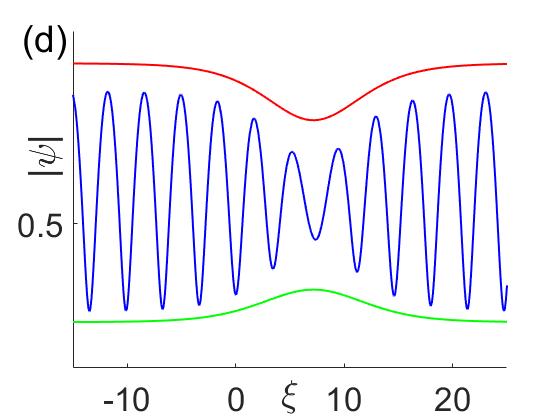}
       \caption{(a): Envelope diagram of the CD-type elliptic dark soliton solution with parameters $k=1/2$, $\epsilon_1=\epsilon_2=1$, $l=-\ii\tau/20$, $a_1=3\ii$, $z_1=2\ii/5+\ii\tau/4$. (b): Profile of the CD-type elliptic dark soliton at $\eta=0$ for the same set of parameters in (a). (c): Envelope diagram of the CU-type elliptic dark soliton solution with parameters $k=1/2$, $\epsilon_1=\epsilon_2=1$, $l=-\ii\tau/20$, $a_1=3\ii$, $z_1=-\ii/6-\ii\tau/4$. (d): Profile of the CU-type elliptic dark soliton at $\eta=0$ for the same set of parameters in (c).}
 \label{fig5}
\end{figure}

 Based on the explicit expressions of the supremum and infimum for $\psi^{[1]}$ given in Eqs. \eqref{inf1}, \eqref{inf2} and \eqref{sup}, we could provide the envelope diagrams for two types of single elliptic dark soliton solutions shown in Fig. \ref{fig5}.
 Among them, the CD-type elliptic dark soliton solution shows a concave downward configuration on both the upper and lower envelope surfaces with the parameter $z_1$ satisfying $\Re(z_1)=\ii\tau/4$. 
Plugging $k=1/2$, $l=-\ii\tau/20$, $a_1=3\ii$, $z_1=2\ii/5+\ii\tau/4$ into Eqs. \eqref{psi1}, \eqref{inf1} and \eqref{sup}, we obtain the envelope diagram of the CD-type elliptic dark soliton shown in Fig. \ref{fig5}(a). As for the profile diagram in Fig. \ref{fig5}(b), the red line segment represents the supremum, which can be obtained from Eq. \eqref{sup}, and the blue line segment represents the infimum, which can be obtained from Eq. \eqref{inf1}. 

Meanwhile, the CU-type elliptic dark soliton solution exhibits a concave downward shape on the upper envelope surface and a convex upward shape on the lower envelope surface with $\Re(z_1)=-\ii\tau/4$. When choosing $k=1/2$, $l=-\ii\tau/20$, $a_1=3\ii$, $z_1=-\ii/6-\ii\tau/4$ in Eqs. \eqref{psi1}, \eqref{inf2} and \eqref{sup}, the envelope diagram of the CU-type elliptic dark soliton could be shown in Fig. \ref{fig5}(c). The corresponding profile diagram could be shown in Fig. \ref{fig5}(d) by setting $\eta=0$, where the red line segment and the blue line segment represent the supremum and infimum respectively. Notably, the red line segment and the blue line segment exhibit two distinct structures. The red line segment can be regarded as a dark soliton and the blue line segment can be regarded as a bright soliton, which correspond to the expressions of the supremum and infimum we obtained in Eqs. \eqref{inf2}, \eqref{sup}.

It can be seen from Fig. \ref{fig5} that when $\Re(z_1) = \pm\ii\tau/4$, the CD-type and the CU-type of single elliptic dark soliton solutions show distinct structures on the lower envelope surface, which correspond well to the expressions presented in Eqs. \eqref{inf1} and \eqref{inf2}.  Remarkably, the CD-type elliptic dark soliton in Fig. \ref{fig5}(a) is similar in structure to the single dark soliton shown in \cite{wang2024dark}, while the CU-type elliptic dark soliton shown in Fig. \ref{fig5}(c) has not been reported yet.

Next, we turn our attention to the dynamical behaviors of $\psi^{[1]}$ as $l = 0$. 
When $l=0$ and $z_1\in S$ with $\Re(z_1)=\pm\ii\tau/4$, the infimum of $\psi^{[1]}$ is
    \begin{equation}\label{infsup}
     \begin{split}
        \psi_{\mathrm{inf}}^{[1]}(\xi,\eta)=&\ii\sqrt{\hat{d}_m}\alpha\frac{\vartheta_2}{\vartheta_3^2}\mathrm{sech}(\hat{L}-\rho_1)\ee^{\omega_2\eta},
     \end{split}
    \end{equation}
   and the supremum is defined in Eq. \eqref{sup} with $l=0$, where
     $\hat{d}_m=\mathrm{max}[0,d_m]$ and $\hat{L}$, $\rho_1$, $d_m$ are defined in Eq. \eqref{inf1}. 
According to the above equations, we find that when $l = 0$, there is only one expression for the supremum and infimum with respect to different values of $z_1$.
In this case, it demonstrates a CU-type elliptic dark soliton solution with $z_1\in S$.
  
By deriving the expressions for the supremum and infimum, we can intuitively analyze two dynamical behaviors and such dynamical behaviors are closely related to the values of $z_1$ and $l$. Since the value of $\lambda$ depends on the values of $z_1$ and $l$,
in what follows, we intend to study the corresponding relationship between different values of $\lambda$ and different dynamic behaviors of the single elliptic dark solutions. By analyzing the values of $\lambda$  corresponding to different $z$ and $l$, we arrive at the following conclusions.
When $l\neq 0$, for any $k\in (0,1)$, $z=\ii z_c+\ii\tau/4$ with $z_c\in\mathbb{R}$ such that $\lambda(z)\in (\hat{\lambda}_1,\hat{\lambda}_2)$, $\psi^{[1]}$ demonstrates a CD-type elliptic dark soliton solution.
When $z=\ii z_c-\ii\tau/4$ with $z_c\in\mathbb{R}$ such that $\lambda(z)\in (\hat{\lambda}_3,\hat{\lambda}_4)$, $\psi^{[1]}$ demonstrates a CU-type elliptic dark soliton solution,
where $\hat{\lambda}_{1,2,3,4}$ are defined as $\hat{\lambda}_{1,4}=\mp\sqrt{2}\alpha(k\cn(4\ii lK)\mp\ii k\sn(4\ii lK)+\dn(4\ii lK))/2$ and $\hat{\lambda}_{2,3}=\hat{\lambda}_{1,4}\pm\sqrt{2}\alpha\dn(4\ii lK)-\sqrt{2}\ii\alpha k^2\cn(4\ii lK)\sn(4\ii lK)/\dn(4\ii lK)$.
    Specifically, when $l=0$, $\hat{\lambda}_{1,2,3,4}$ could be written as            $\hat{\lambda}_{1,4}=\mp \sqrt{2}\alpha(k+1)/2$, $
            \hat{\lambda}_{2,3}=\mp \sqrt{2}\alpha(-k+1)/2$.
    $\psi^{[1]}$ with $\lambda(z)\in (\hat{\lambda}_1,\hat{\lambda}_2)\cup (\hat{\lambda}_3,\hat{\lambda}_4)$ demonstrates a CU-type elliptic dark soliton solution.
    In particular, when $z_1=\pm\ii\tau/4$ or $\pm1/2\pm\ii\tau/4$ (i.e. $\lambda=\hat{\lambda}_{1,2,3,4}$), from Eq. \eqref{psi1} we obtain that the exact expression of $\psi^{[1]}$ is equal to $\psi$ in Eq. \eqref{solu2}. Therefore, we could conclude that $\psi^{[1]}$ degenerates into the background solution at points $\hat{\lambda}_{1,2,3,4}$. 
    
Based on the above analysis, we present a more intuitive diagram to illustrate the relationship between the values of $z$ corresponding to different values of $\lambda(z)$ and the dynamic behaviors. Relying on the properties of Jacobi elliptic functions, it could be verified that $\lambda(\ii z_c\pm\ii\tau/4)=\lambda(\ii(z_c\pm n)\pm\ii\tau/4)$, $n\in\mathbb{N}^{+}$. That is to say, the period of $\lambda(\ii z_c\pm\ii\tau/4)$ with respect to $z_c$ is $1$. Hence, we only show one period in the following figures.
\begin{figure}[htp]
	\centering
        \includegraphics[width=1\linewidth,trim=100 0 60 10,clip]{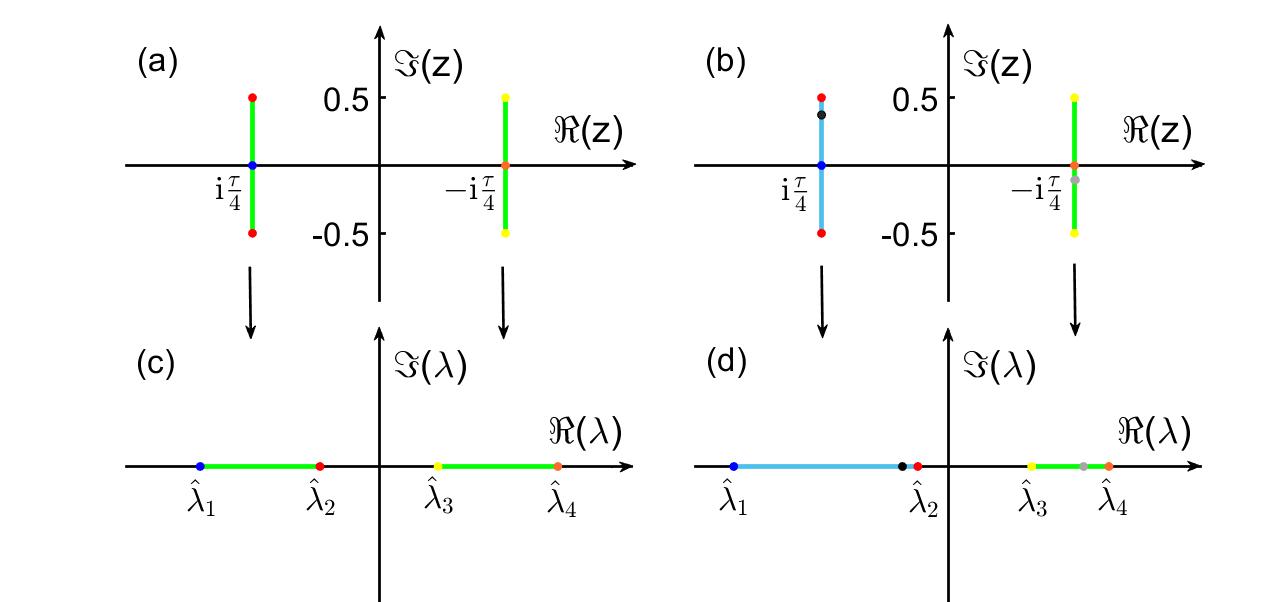}
	\caption{Relationship of $z$ and $\lambda$ with $k=1/2$, $\epsilon_1=\epsilon_2=1$. (a) and (c): $z$-plane and the corresponding $\lambda$-plane with $l=0$.  (b) and (d): $z$-plane and the corresponding $\lambda$-plane with $l=-\ii\tau/20$, the black point and gray point correspond to the data points of the single elliptic-dark soliton shown in Fig. \ref{fig5}(a) and \ref{fig5}(c) respectively.}
 \label{fig3}
\end{figure}

 The relationship between $\lambda$ and the types of the single elliptic dark soliton solutions is shown in Fig. \ref{fig3}. The points of different colors in the $z$-plane correspond to those of the same colors in the $\lambda$-plane. For example, the red points in Fig. \ref{fig3}(a) are mapped to the red point in Fig. \ref{fig3}(c).
 When $\lambda$ is located on the green line segments in Fig. \ref{fig3}(c) and \ref{fig3}(d), the single elliptic dark soliton is a CU-type elliptic dark soliton, which corresponds to Eqs. \eqref{infsup} and \eqref{inf2} respectively.
 As for the blue line segment in Fig. \ref{fig3}(d), the single elliptic-dark soliton shows a CD-type elliptic dark soliton, which is consistent with Eq. \eqref{inf1}. 
Additionally, when $\lambda$ is located on the endpoints at both ends of the blue and green line segments in Fig. \ref{fig3}(b) and \ref{fig3}(d), the solution $\psi^{[1]}$ is similar to the background solutions on both the dynamic behaviors and the exact expressions. 

To conclude, by choosing different parameters $z_1$, we obtain different dynamical behaviors and summarize them as follows.
\begin{itemize}
    \item[Case 1]: When $l\neq 0$, if $z_1$ satisfies $\Re(z_1)=\ii\tau/4$ and $\Im(z_1)\in(-0.5,0.5)$, $\psi^{[1]}$ demonstrates a CD-type elliptic soliton solution (see Fig. \ref{fig5}(a)).
    \item[Case 2]: When $l\neq 0$, if $z_1$ satisfies $\Re(z_1)=-\ii\tau/4$ and $\Im(z_1)\in(-0.5,0.5)$, $\psi^{[1]}$ demonstrates a CU-type elliptic soliton solution (see Fig. \ref{fig5}(c)). 
     \item[Case 3]: When $l=0$, if $z_1$ satisfies $\Re(z_1)=\pm \ii\tau/4$ and $\Im(z_1)\in(-0.5,0.5)$, $\psi^{[1]}$ demonstrates a CU-type elliptic soliton solution, corresponding to Eq. \eqref{infsup}. 
    \item[Case 4]: When $z_1=\pm \ii\tau/4$ or $\pm\ii/2\pm\ii\tau/4$, $\psi^{[1]}$ degenerates to the periodic background solution.
\end{itemize}

In this section, we mainly focus on two types of single dark soliton solutions for the Hirota equation \eqref{hirota} and their dynamical behaviors. Additionally, we present the expression of these solutions in terms of the theta functions. In what follows, we aim to construct multi-elliptic dark soliton solutions by the Darboux-B\"{a}cklund transformation.

\section{Collision property of multi-elliptic dark soliton solutions}\label{sec3}

To construct the multi-elliptic dark soliton solutions, we consider the multifold Darboux matrix. 
After $N$ times of the iteration, the $N$-fold Darboux matrix could be expressed as $\mathbf{T}^{[N]}=\mathbf{T}_N\mathbf{T}_{N-1}\cdots\mathbf{T}_1=\mathbb{I}-\mathbf{X}_N\mathbf{M}_N^{-1}\mathbf{D}_N^{-1}\mathbf{X}_N^{\dagger}$, where $\mathbf{X}_N=[\Phi_1,\Phi_2,\cdots,\Phi_n]$, $\Phi_i=\Phi(\xi,\eta;\lambda_i)\mathbf{c}_i$, $\mathbf{c}_i=[1,c_{i2}]^{\top}$, $c_{i2}=(\lambda_i-\lambda_i^{*})a_i$, $\mathbf{D}_N=\mathrm{diag}(\lambda-\lambda_1,\lambda-\lambda_2,\cdots,\lambda-\lambda_N)$ and the $(i,j)$-element of the matrix $M_N$ is $\Phi_i^{\dagger}\sigma_3\Phi_j/(2\lambda_j-2\lambda_i^{*})$.
Based on them, it is easy to obtain the multi-elliptic dark soliton solutions \cite{feng2020multi}:
\begin{equation}\label{darb}
    \!\!\psi^{[N]}=\psi^{[N]}(\xi,\eta)=\frac{\psi^{1-N}\mathrm{det}(\psi\mathbf{M}_N-\ii \mathbf{X}^{\dagger}_{N,2}\mathbf{X}_{N,1})}{\mathrm{det}(\mathbf{M}_N)},\!\!
\end{equation}
where $\mathbf{X}_{N,i}$ is the $i$-row of the matrix $\mathbf{X}_N$, and $\psi$ is given by Eq. \eqref{solu2}.
When $z_i\in S$ (defined in Eq. \eqref{f16}) satisfies $\Re(z_i)=\pm\ii\tau/4$, the multi-elliptic dark soliton solutions of the Hirota equation \eqref{hirota} can be represented as the following determinant form:
    \begin{equation}\label{eq:psi-solution}
      \!\!\!\!\!  \psi^{[N]}\!\!=\!\chi
        \frac{\vartheta_1\!\left(\hat{\alpha}\xi +2\ii l  \right)\mathrm{det}(\mathcal{E}^{\dagger}\mathcal{P}^{\dagger}\mathcal{H}\mathcal{P}^{\!-1}\mathcal{E}\!+\!\mathbb{I}_N)}{\vartheta_4\left(\hat{\alpha}\xi \right)\mathrm{det}(\mathcal{E}^{\dagger}\mathcal{G}\mathcal{E}+\mathbb{I}_N)}\ee^{\omega_1\xi+\omega_2\eta},
    \end{equation}
    under the transformation $(\xi,\eta)$ defined in Eq. \eqref{eq:transformation}, where $\mathcal{E}=\mathrm{diag}(E_1(z_1),E_1(z_2),\cdots,E_1(z_N))$, $\mathcal{P}=\diag(p_1,p_2\cdots,p_N)$ with $p_i=\vartheta_4(\ii(z_i-l) )/\vartheta_1(-\ii(z_i+l) )$, and the $(i,j)$-element of matrices $\mathcal{G}$ and $\mathcal{H}$ are
    \begin{equation}\nonumber
    \begin{split}
        (\mathcal{G})_{ij}&=\frac{\vartheta_4\left(\hat{\alpha}\xi +\ii(z_i^{*}-z_j) \right)}{2\hat{a}_i\vartheta_1(\ii(z_i^{*}-z_j) )\vartheta_4\left(\hat{\alpha}\xi \right)},\\
        (\mathcal{H})_{ij}&= \frac{ \vartheta_1\left(\hat{\alpha}\xi +\ii(z_i^{*}-z_j) +2\ii l \right)}{-2\hat{a}_i\vartheta_1(\ii(z_i^{*}-z_j) )\vartheta_1\left(\hat{\alpha}\xi +2\ii l  \right)},
        \end{split}
    \end{equation}
    with $\hat{a}_i=\mathrm{Im}(a_i)\ee^{-\ii\tau /4}>0$. It could be verified that $\psi^{[N]}$ is analytic for all $(\xi,\eta)\in \mathbb{R}^2$ with the fact that $\mathrm{det}(\mathcal{E}^{\dagger}\mathcal{G}\mathcal{E}+\mathbb{I}_N)>0$, one can refer to Appendix \ref{app3} for detailed proof.
    
\begin{figure}[htp]
	\centering
    \includegraphics[width=0.43\linewidth]{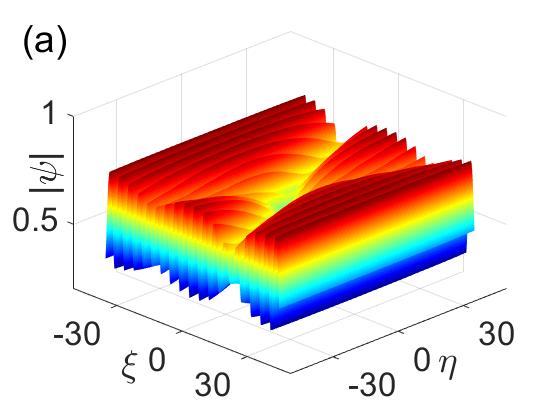}
    \includegraphics[width=0.43\linewidth]{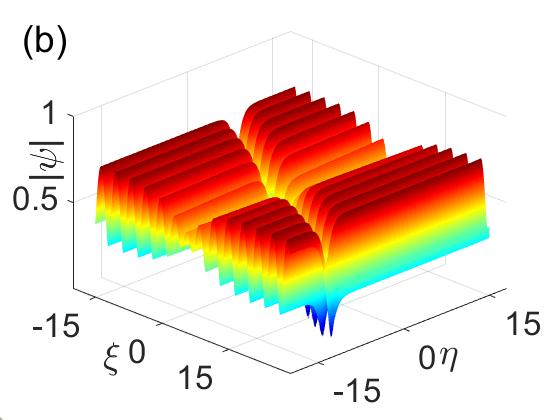}\\
     \includegraphics[width=0.43\linewidth]{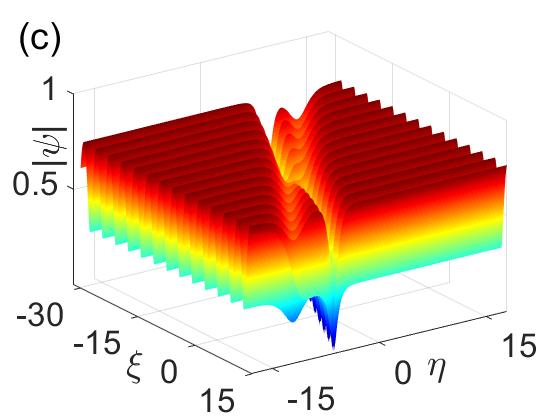}
    \includegraphics[width=0.43\linewidth]{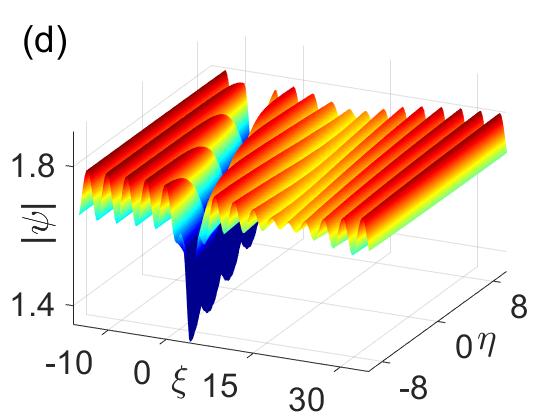}
	\caption{Two-elliptic dark soliton solutions with $k=1/2$, $\epsilon_1=\epsilon_2=1$, $\alpha=1$, $a_1=a_2=3\ii$, $l=-\ii\tau/20$, $z_1=-\ii/6-\ii\tau/4$, $z_2=-\ii/3-\ii\tau/4$ in (a),  $z_1=-\ii/4-\ii\tau/4$, $z_2=2\ii/7+\ii\tau/4$ in (b), $z_1=\ii/3+\ii\tau/4$, $z_2=\ii/4+\ii\tau/4$ in (c) and $l=-13\ii\tau/80$, $z_1=13\ii/50+\ii\tau/4$, $z_2=-13\ii/50-\ii\tau/4$ in (d).}
 \label{fig1}
\end{figure}

Through choosing different parameters $z_i\in S$, $i=1,2,\cdots,N$, we could obtain $N$-elliptic dark soliton solutions. The dynamic behaviors of these $N$ dark solitons are determined by the parameters $z_i$ and $l$ we select.
Selecting a set of parameters based on Eq. \eqref{eq:psi-solution} and taking $N=2$, we are able to illustrate $\psi^{[2]}$ shown in Fig. \ref{fig1}.
In particular, by choosing the parameters $\alpha=1$, $k=1/2$, $l=-\ii\tau/20$, $\epsilon_1=\epsilon_2=1$, $a_1=a_2=3\ii$, $z_1=-\ii/6-\ii\tau/4$, $z_2=-\ii/3-\ii\tau/4$, we can obtain Fig. \ref{fig1}(a). It demonstrates the interaction between two elliptic dark soliton solutions, both of which are CU-type elliptic dark solitons. 
Meanwhile, we get Fig. \ref{fig1}(b) by setting $z_1=-\ii/4-\ii\tau/4$, $z_2=2\ii/7+\ii\tau/4$. This figure shows the interaction between elliptic dark solitons exhibiting two different types, where one is CU-type elliptic dark soliton and the other is CD-type elliptic dark soliton. 
Changing the parameters $z_1=\ii/3+\ii\tau/4$, $z_2=\ii/4+\ii\tau/4$, we provide Fig. \ref{fig1}(c) which shows the interaction between two CD-type elliptic dark solitons. It can be noted that when we select the parameters $l=-13\ii\tau/80$, $z_1=13\ii/50+\ii\tau/4$, $z_2=-13\ii/50-\ii\tau/4$, the two-elliptic dark soliton solution exhibits two parallel dark solitons, one of which is a CD-type elliptic dark soliton and the other is a CU-type elliptic dark soliton.

As can be seen from Fig. \ref{fig1}, the two-elliptic dark soliton solutions shown in Figs. \ref{fig1}(a)–(c) demonstrate collisions between two dark solitons. In contrast, in Fig. \ref{fig1}(d), no such collisions take place. An important step in analyzing whether collisions occur is to analyze whether the velocities of the two solitons are the same. We define the velocity of each dark soliton as $\nu_i=\Re(\tilde{V}_i)/\Re(\tilde{W}_i)$ with $\tilde{W}_i=W_1(z_i)$ and $\tilde{V}_i=V_1(z_i)$.

\begin{figure}[htp]
		\centering
		\includegraphics[width=0.43\linewidth]{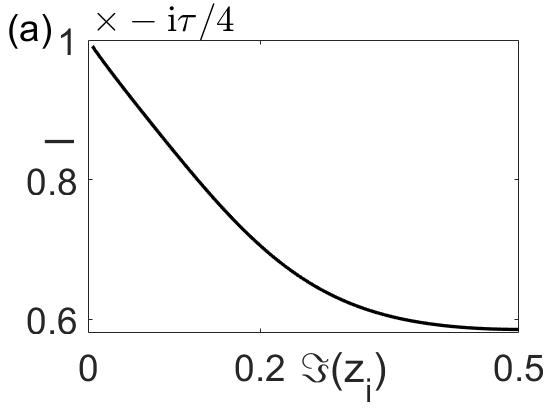}
            \includegraphics[width=0.43\linewidth]{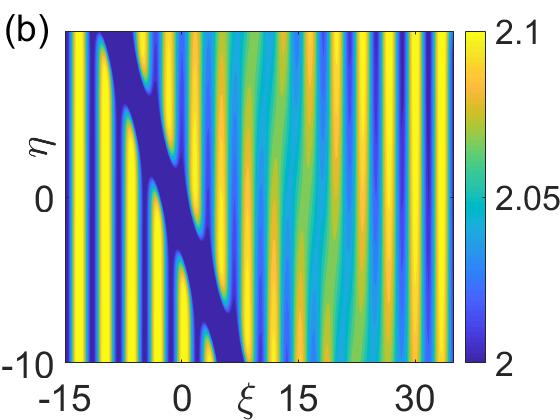}
		\caption{(a) The relationship between $l$ and $\Im(z_i)$ when $\nu_1=\nu_2$ with $k=1/2$, $\alpha=1$. (b) The density plot of two-elliptic dark soliton solution with $ \nu_1=\nu_2\approx 0.77$. The parameters are chosen as $k=1/2$, $\alpha=1$, $\epsilon_1=\epsilon_2=1$, $l=-7\ii\tau/40$, $a_1=a_2=3\ii$, $z_1=20.48\ii+\ii\tau/4$, $z_2=20.48\ii-\ii\tau/4$.}
		\label{fig:samev}
	\end{figure}

Consider the two-elliptic dark soliton solutions that include a CD-type elliptic dark soliton and a CU-type elliptic dark soliton with velocities $\nu_1$ and $\nu_2$ respectively. 
Based on the definition of velocity, we substitute $k=1/2$, $\alpha=1$ into the equation $\nu_1=\nu_2$ and analyze the relationships between the velocities of solitons and the parameters $z_i$ and $l$. Then we obtain Fig. \ref{fig:samev}(a) which reveals that when $l\in [-59\ii\tau/400,-\ii\tau/4) $, such a two-elliptic dark soliton solution with identical velocities can be presented.
By setting the parameters $k=1/2$, $\alpha=1$, $\epsilon_1=\epsilon_2=1$, $l=-7\ii\tau/40$, $a_1=a_2=3\ii$, $z_1=128\ii/625+\ii\tau/4$ and $z_2=-128\ii/625-\ii\tau/4$ , we are able to obtain Fig. \ref{fig:samev}(b), which shows that a CU-type elliptic dark soliton and a CD-type elliptic dark soliton move at the same speed $\nu_1=\nu_2\approx 0.77$. Since the velocities of the two dark solitons are identical, they will maintain a fixed relative position during propagation, and thus they will propagate without colliding with each other.
	
\begin{figure}[htp]
		\centering
		\includegraphics[width=1\linewidth,trim=100 0 60 10,clip]{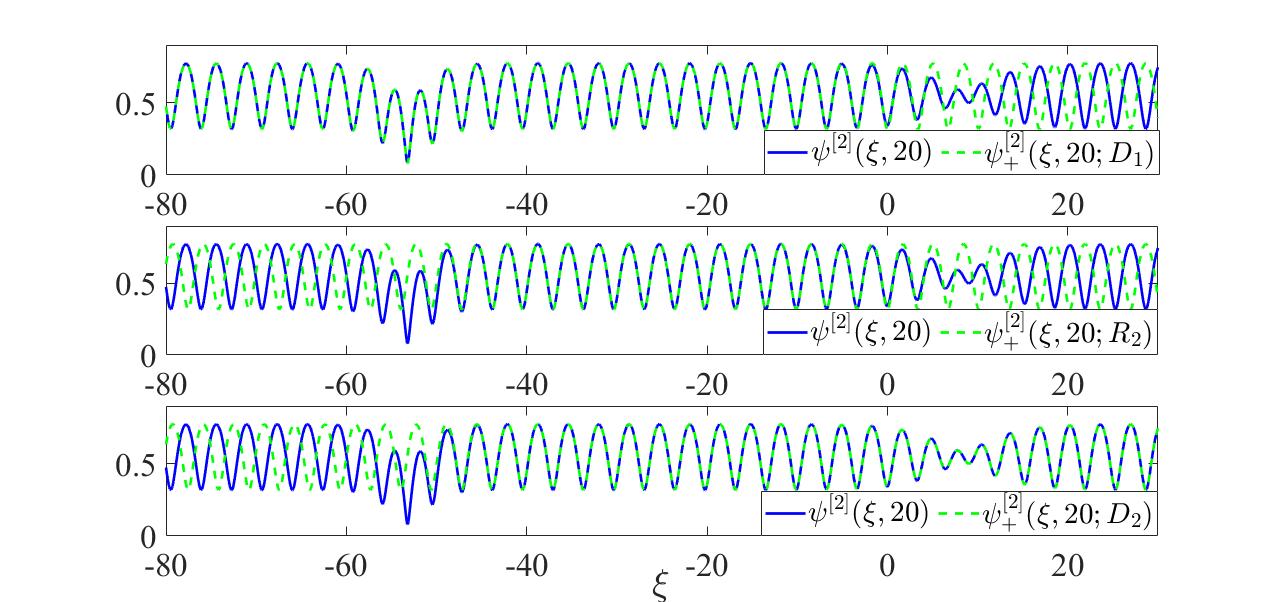}
		\caption{The asymptotic dynamic behaviors of the solution $\psi^{[2]}$ along the trajectories $D_{1,2}$ and on the region $R_2$ at the time $\eta=20$ with $k=1/2$, $\alpha=1$, $\epsilon_1=\epsilon_2=1$, $l=-\ii\tau/20$, $a_1=a_2=3\ii$, $z_1=-\ii/4-\ii\tau/4$, $z_2=2\ii/7+\ii\tau/4$.}
		\label{fig:asyr}
	\end{figure}

When the velocities of the solitons are inconsistent, collisions will occur. We proceed to consider the dynamics of solutions $\psi^{[N]}$ along the trajectories of the elliptic dark solitons. 
For ease of representation, we introduce the lines $D_i$ as $ D_i:=\Re(\tilde{W}_i)\left(\xi-\nu_i \eta\right)$.
Without loss of generality, we set $\Re(\tilde{W}_i)>0$, $i=1,2,\cdots,N$, and $\nu_1< \nu_2< \cdots < \nu_N$. 
Based on the above assumption, as $\eta\rightarrow \pm \infty$, the asymptotic expressions of the multi-elliptic dark soliton solution $\psi^{[N]}$ along the trajectories $D_i$, $i=1,2,\cdots,N$ could be expressed as: 
\begin{equation}\label{f50}
\psi^{[N]}_{D_i^{\pm}}\rightarrow(-1)^i\chi \mathbf{r}^{\pm}_i\frac{H^{\pm}\ee^{2 D_i+\gamma_i^{\pm}}-1}{G^{\pm}\ee^{2 D_i}+1}\ee^{\omega_1\xi+\omega_2\eta},
\end{equation}
where $\psi^{[N]}_{D_i^{\pm}}=\psi^{[N]}_{\pm}(\xi,\eta;D_i)$, 
$H^{\pm}=\vartheta_1(\hat{\alpha}\xi +\mathbf{s}_{i+1}^{\pm}+2\ii l )/(2\hat{a}_i\vartheta_1(\hat{\alpha}\xi +\mathbf{s}_i^{\pm}+2\ii l )\vartheta_1(\ii (z_i^*-z_i) ))$, 
$G^{\pm}=\vartheta_4(\hat{\alpha}\xi +\mathbf{s}_{i+1}^{\pm})/(2\hat{a}_i\vartheta_4(\hat{\alpha}\xi +\mathbf{s}_i^{\pm})\vartheta_1(\ii (z_i^*-z_i) ))$,
$\mathbf{s}_i^{+}=\sum_{j=1}^{i-1}\ii(z_j^*-z_j) $, $\mathbf{s}_i^{-}=\sum_{j=i+1}^{N}\ii(z_j^*-z_j) $, $\mathbf{r}_i^{+}=\prod_{m=1}^{i-1}p_m^*/p_m$, $\mathbf{r}_i^{-}=\prod_{m=i+1}^{N}p_m^*/p_m$and $\gamma_i^{\pm}$ is defined as
\begin{equation}\label{eq:s-r-gamma-define}
\begin{split}
\gamma_i^{+}=&\ln\left(\prod_{m=1}^{i-1}\frac{p_i^*p_i^{-1}}{\vartheta_1(\ii (z_m^*-z_i) )\vartheta_1(\ii (z_i^*-z_m) )}\right),\\
\gamma_i^{-}=&\ln\left(\prod_{m=i+1}^{N}\frac{p_i^*p_i^{-1}}{\vartheta_1(\ii (z_m^*-z_i) )\vartheta_1(\ii (z_i^*-z_m) )}\right).
\end{split}
\end{equation}
From Eq. \eqref{f50}, $\psi^{[N]}_{D_i^{\pm}}$ could be considered as a shift on the solution $\psi^{[1]}$ in Eq. \eqref{psi1} and the amplitude is multiplied by a constant $(-1)^{i-1}\mathbf{r}_i^{\pm}\ee^{-\omega_1\mathbf{s}_{i}^{\pm}/\hat{\alpha}}$ with the shift being $\mathbf{s}_i^{\pm}/\hat{\alpha}$.
As $\eta \rightarrow\pm \infty$, the asymptotic expressions of the multi-elliptic dark soliton solution $\psi^{[N]}$ in the region $R_i$ between the line $D_{i-1}$ and $D_i$ could be expressed as 
\begin{equation}\label{eq:R}
        \begin{split}
			\psi^{[N]}_{R_i^{\pm}}\rightarrow
			(-1)^i\chi\mathbf{r}_i^{\pm}\frac{\vartheta_1(\hat{\alpha}\xi +\mathbf{s}_i^{\pm}+2\ii l )}
			{\vartheta_4(\hat{\alpha}\xi +\mathbf{s}_i^{\pm})}\ee^{\omega_1\xi+\omega_2\eta},
        \end{split}
\end{equation}
where $\psi^{[N]}_{R_i^{\pm}}=\psi^{[N]}_{\pm}(\xi,\eta;R_i)$. Considering the asymptotic dynamic behaviors of two-elliptic dark soliton solution $\psi^{[2]}$, we exhibit Fig. \ref{fig:asyr} with parameters $k=1/2$, $\epsilon_1=\epsilon_2=1$, $l=-\ii\tau/20$, $a_1=a_2=3\ii$, $z_1=-\ii/4-\ii\tau/4$, $z_2=2\ii/7+\ii\tau/4$. 
By substituting the above parameters into Eqs. \eqref{eq:psi-solution}, \eqref{f50} and \eqref{eq:R}, we can obtain the plot of $\psi^{[2]}(\xi,20)$, $\psi_{+}^{[2]}(\xi,20;D_{1,2})$ and $\psi_{+}^{[2]}(\xi,20;R_2)$ respectively.
In Fig. \ref{fig:asyr}, the blue curves indicate the solution $\psi^{[2]}(\xi,20)$ and the green dashed curves represent the functions $\psi_{+}^{[2]}(\xi,20;D_1)$, $\psi_{+}^{[2]}(\xi,20;R_2)$ and $\psi_{+}^{[2]}(\xi,20;D_2)$ from the top to bottom. Upon careful observation and comparison of these curves, we notice that near the trajectories $D_{1,2}$ and within the region $R_2$, the asymptotic dynamical behavior depicted in green perfectly matches the blue curves. 
When time approaches infinity, the dynamical behaviors of the solutions remain unchanged. The CU-type elliptic dark soliton solution remains a CU-type elliptic dark soliton solution, and the CD-type elliptic dark soliton solution also remains a CD-type elliptic dark soliton solution. 
This implies an elastic collision, that is, before and after the collision the dynamical behaviors of the solitons do not change, which respectively correspond to Eq. \eqref{f50} with $\eta \rightarrow \mp \infty$.

\section{Conclusions}\label{conclusion}
In this work, we report CD-type and CU-type elliptic dark soliton solutions for the Hirota equation and reveal the existence conditions of the above two types, which, to the best of our knowledge, have not been previously reported.
The existence conditions for the above two types of solutions are studied by deriving the supremum and infimum of the related solutions.
Furthermore, using the Darboux-B\"{a}cklund transformation, we construct multi-elliptic dark soliton solutions. Two-elliptic dark soliton solutions with the same velocity but different shapes are obtained.
Through asymptotic analysis, the elastic collisions among the above types elliptic dark soliton solutions are revealed.  

This research contributes to a profound understanding of the phenomena associated with dark soliton solutions in the realm of complex dynamics \cite{PhysRevE.106.044214, PhysRevE.109.014201,PhysRevLett.86.2918,PhysRevA.84.041605}.
The existence conditions for CD-type and CU-type elliptic dark soliton solutions provide support for defining the observational conditions of two types of single dark soliton on elliptic function backgrounds in physical experiments.
Additionally, CD-type and CU-type elliptic dark solitons with identical velocities may have applications in experimentally observing the bound states solitons in optic systems \cite{PhysRevLett.121.023905,PhysRevLett.118.243901}. The  elastic collisions phenomena revealed in this study will be observed in experimental physics, such as hydrodynamic and optical experiments \cite{ PhysRevLett.101.130401,PhysRevA.92.053850,PhysRevE.94.012205,PhysRevA.103.013521,PhysRevE.97.032218}.

\begin{acknowledgments}
\vspace{-10pt}
We sincerely thank Professor Liming Ling for his guidance and valuable suggestions to improve this project.
Meanwhile, we sincerely thank Professor Yanhong Qin for her help and valuable discussions.
%The Fundamental research funds for the central universities with contract numbers
% 2232022G-13, 2232023G-13 and 2232024G-13.
\end{acknowledgments}
%\vspace{-10pt}
%\section*{AUTHOR DECLARATIONS}
%\vspace{-10pt}
%\section*{Conflict of Interest}
%\vspace{-10pt}
%The authors have no conflicts to disclose.
%\vspace{-10pt}
%\section*{Author Contributions}
%\vspace{-10pt}
%\noindent\textbf{Qiaofeng Huang}: Conceptualization (equal); Formal analysis (equal); Investigation (equal); Methodology (equal); Software (equal); Validation (equal); Visualization (equal); Writing – original draft
%(equal); Writing – review \& editing (equal).
%\textbf{Xuan Sun}: Funding acquisition (equal); Conceptualization (equal); Methodology (equal); Project administration (equal); Supervision (equal); Validation (equal); Writing – review \& editing (equal).
%\vspace{-10pt}
%\section*{DATA AVAILABILITY}
%\vspace{-10pt}
%Data sharing is not applicable to this article as no new data were created or analyzed in this study.

\appendix
\titleformat{\section}[display]
{\centering \bfseries}{ }{11pt}{ }

\setcounter{section}{0}

\renewcommand{\appendixname}{Appendix\,\,\Alph{section}}
\vspace{-10pt}
\section{\appendixname. The definition of elliptic functions}
\begin{definition}
    The Jacobi elliptic functions are defined as \cite{gradshteyn2014table}:
    \begin{equation}
    \begin{split}
        &\sn(u)=\sn(u,k)=u^{-1}(y,k), \\ &u(y,k)=\int_0^{y}\frac{\dd t}{\sqrt{(1-t^2)(1-k^2t^2)}},
    \end{split}
    \end{equation}
functions $\cn(u,k)$, $\dn(u,k)$ are defined as $\sqrt{1-y^2}$, $\sqrt{1-k^2y^2}$ respectively and $k$ is the modulus. 
\end{definition}
\begin{definition}\label{theta fun}
    The theta functions are defined as the summation \cite{gradshteyn2014table}: 
    \begin{equation}
        \begin{aligned}
           & \vartheta_1(u)\!=\!-\ii\!\!\sum_{n=-\infty}^{\infty}\!(-1)^nq^{(n+\frac{1}{2})^2}\ee^{\ii(2n+1)u\pi},\quad \!\!\vartheta_3(u)\!=\!\!\sum_{n=-\infty}^{\infty}\!q^{n^2}\ee^{\ii2nu\pi},\\
           & \vartheta_2(u)\!=\!\!\sum_{n=-\infty}^{\infty}\!q^{(n+\frac{1}{2})^2}\ee^{\ii(2n+1)u\pi}, \quad \!\!\vartheta_4(u)\!=\!\!\sum_{n=-\infty}^{\infty}(-1)^n \!q^{n^2}\ee^{\ii2nu\pi},\\
        \end{aligned}
    \end{equation}
    where $\vartheta_i(u)=\vartheta_i(u,q)$, $\tau=\ii K^{\prime}/K$ and $q=\ee^{\ii \tau \pi}$ is called the nome of the theta functions.
\end{definition}

\begin{widetext}
\section{\appendixname. Proofs of Eqs. \eqref{f9}, \eqref{solu2} and \eqref{psi}}\label{app1}

 \textbf{The proof of the Eq. \eqref{f9}:}
     From Eq. \eqref{f1} and the definition of $\mu$, we have $v_x=-\sqrt{2}\ii (\mu-\mu^*)v=\sqrt{2R(v)}$, $v_t=-\epsilon_2 s_2v_x$.
     The function $v$ shows periodic dynamic behaviors solely when the equation $R(v) = 0$ has three real roots, denoted as $v_{1,2,3}$. When $l\in[0,-\ii\tau/4)$, $\tau=\ii K^{\prime} /K$, $k\in(0,1)$ and $\alpha>0$, we have 
$0\leq v_1\leq v_2\leq v_3$ and $v$ can oscillate between $v_1$ and $v_2$. Furthermore, we can deduce that there exists a one-to-one correspondence between the triple tuples $(v_1,v_2,v_3)$ and $(\alpha,k,l)$, which can be referred to \cite{ling2023stability} for detailed proof. 
Combining with properties of Jacobi elliptic function, we can obtain
\begin{equation}\label{f5}
     R(v)=(v-v_1)(v-v_2)(v-v_3)=8k^4\alpha^6\sn^2(\alpha (x-\epsilon_2 s_2t))\cn^2(\alpha (x-\epsilon_2 s_2t))\dn^2(\alpha (x-\epsilon_2 s_2t)).
 \end{equation}
     On the other hand, we have
     $v_x=4k^2\alpha^3 \sn(\alpha (x-\epsilon_2 s_2t))\cn(\alpha (x-\epsilon_2 s_2t))\dn(\alpha (x-\epsilon_2 s_2t))$,
     which implies $v_x^2=2R(v)$. Similarly, we can obtain $v_t^2=2\epsilon_2^2 s_2^2R(v)$ and thus the proof is completed.

\textbf{The proof of the Eq. \eqref{solu2}:}
 From Eqs. \eqref{f1} and \eqref{f2}, we know $(\ln\psi)_x=-\sqrt{2}\ii \mu=-\sqrt{2}\ii(s_3+\ii\sqrt{R(v)})/(2v)$, $(\ln\psi)_t=-\epsilon_2s_2 (\ln\psi)_x+\ii\epsilon_1 s_2-\sqrt{2}\ii \epsilon_2s_3$.
 Integrating the above equations under the transformation between $(\xi,\eta)$ and $(x,t)$ defined in Eq. \eqref{eq:transformation},
 we could obtain $\psi(\xi,\eta)=\sqrt{v(\xi)} \exp(-\int_0^{\xi}\sqrt{2}\ii s_3/(2v(z))\dd z+\ii(\epsilon_1s_2-\sqrt{2}\epsilon_2s_3)\eta)$.
According to the representation of $R(v)$ in Eqs. \eqref{f2} and \eqref{f5}, we can obtain $s_2=-(v_1+v_2+v_3)/2$ and $s_3=\sqrt{v_1v_2v_3}$. 
Subsequently, by using the formula (\cite{feng2020multi} p.11),
we obtain $\int_0^{\xi}s_3/v(z)\dd z=-\sqrt{2}\ii \int_0^{\alpha\xi}\sn(4\ii l K)\cn(4\ii l K)\dn(4\ii l K)/(\sn^2(z)-\sn^2(4\ii l K))\dd z=-\sqrt{2}\ii (\ln \sqrt{\vartheta_1(2\ii l  -\hat{\alpha}\xi )/\vartheta_1(2\ii l  +\hat{\alpha}\xi )}+\alpha Z(4\ii l K)\xi).$

On the other hand, based on addition formulas for the theta functions and the relationship between theta functions and elliptic functions (\cite{gradshteyn2014table} p.888), we have $v(\xi)=2\alpha^2 k^2\left(\sn^2(\alpha\xi)-\sn^2(4\ii l K)\right)
       =-2\alpha^2\vartheta_2^2\vartheta_4^2\vartheta_1\left(\hat{\alpha}\xi +2\ii l \right)\vartheta_1\left(2\ii l -\hat{\alpha}\xi \right)/(\vartheta_3^2\vartheta_4^2(2\ii l )\vartheta_4^2\left(\hat{\alpha}\xi \right))$,
    where $\hat{\alpha}$ is defined in Eq. \eqref{solu2}. Above all, the solution $\psi(\xi,\eta)$ can be turned into the form of theta functions \eqref{solu2} and the proof is done.

\textbf{The proof of the Eq. \eqref{psi}:}
    According to the Lax pair \eqref{lax}, under the $(\xi,\eta)$ coordinate, we can derive $\Phi_{1i}\equiv\Phi_{1i}(\xi,\eta;\lambda)$, $i=1,2$ through the following equations:
     \begin{equation}\label{f6}
        \begin{aligned}
            \Phi_{1i,\xi}=&\mathbf{U}_{11}\Phi_{1i} - \frac{2\ii\psi^{*}(\lambda-\mu^{*})\mathbf{U}_{12}\Phi_{1i}}{v+2\lambda^2+s_2+(-1)^{i-1}2y},\quad
            \Phi_{1i,\eta}=\hat{\mathbf{V}}_{11}\Phi_{1i}+\frac{v+2\lambda^2+s_2+(-1)^i 2y}{2\ii\psi(\lambda-\mu)}\hat{\mathbf{V}}_{12}\Phi_{1i},
        \end{aligned}
    \end{equation}
where $\mathbf{U}_{ij}$, $\hat{\mathbf{V}}_{ij}$ are the $(i,j)$-element of $\mathbf{U}(\lambda;\mathbf{Q})$ and $\hat{\mathbf{V}}(\lambda;\mathbf{Q})=\mathbf{V}+\epsilon_2 s_2\mathbf{U}$, respectively. Simplifying the above equations, we obtain
   $(\ln \Phi_{1i})_{\xi}=-\sqrt{2}\ii\lambda\xi/2+(\sqrt{2}\ii(2\lambda\beta_i+s_3)+(v+\beta_i)_{\xi})/(2(v+\beta_i))$, $
     (\ln \Phi_{1i})_{\eta}=-\epsilon_2s_2(\ln \Phi_{1i})_{\xi}+\ii\epsilon_1s_2/2-\sqrt{2}\ii\epsilon_2s_3/2+(-1)^{i+1}\ii(\sqrt{2}\epsilon_2\lambda-\epsilon_1)y$,
where $i=1,2$ and
\begin{equation}\label{f7}
    \beta_1=2\lambda^2+s_2+2y,\quad \beta_2=2\lambda^2+s_2-2y.
\end{equation}
Then we obtain $\Phi_{1i}(\xi,\eta;\lambda)=\sqrt{(v(\xi)+\beta_i)/(v(0)+\beta_i)}\ee^{\theta_i}$, where $\Phi_{1i}(0,0;\lambda)=1$, $i=1,2$ and
\begin{equation}\label{f8}
\begin{aligned}
    \theta_i=&-\frac{\sqrt{2}}{2}\ii \lambda \xi+\frac{\sqrt{2}}{2}\ii\int_0^{\xi}\frac{2\lambda\beta_i+s_3}{v(z)+\beta_i}\dd z+\ii\left(\frac{1}{2}\epsilon_1 s_2-\frac{\sqrt{2}}{2} \epsilon_2s_3+(-1)^{i+1}(\sqrt{2}\epsilon_2\lambda-\epsilon_1)y\right)\eta.
\end{aligned}
\end{equation}
Similarly, based on the relation $\Phi_{2i}=r_i\Phi_{1i}$, $i=1,2$, and Eq. \eqref{f6}, we have
\begin{equation}\nonumber
\begin{split}
    \Phi_{21}(\xi,\eta;\lambda)&=\Phi_{21}(0,0;\lambda)\sqrt{\frac{v(\xi)+\beta_2}{v(0)+\beta_2}}\ee^{-\theta_2},\quad
    \Phi_{22}(\xi,\eta;\lambda)=\Phi_{22}(0,0;\lambda)\sqrt{\frac{v(\xi)+\beta_1}{v(0)+\beta_1}}\ee^{-\theta_1},
    \end{split}
\end{equation}
with $\Phi_{21}(0,0;\lambda)=-\sqrt{(v(0)+\beta_2)/(v(0)+\beta_1)}$ and $\Phi_{22}(0,0;\lambda)=-\sqrt{(v(0)+\beta_1)/(v(0)+\beta_2)}$ for better symmetry. 
Then we come up with a fundamental solution of Lax pair \eqref{lax} after ignoring the constant factors of vector solutions:
    \begin{equation}\label{eq:Phi}
        \Phi=\begin{bmatrix}\sqrt{v(\xi)+\beta_1}\ee^{\theta_1}&\sqrt{v(\xi)+\beta_2}\ee^{\theta_2}\\ -\sqrt{v(\xi)+\beta_2}\ee^{-\theta_2}&-\sqrt{v(\xi)+\beta_1}\ee^{-\theta_1}\end{bmatrix},
    \end{equation}
    where $\Phi:=\Phi(\xi,\eta;\lambda)$, the transformation between $(\xi,\eta)$ and $(x,t)$ is defined in Eq. \eqref{eq:transformation}, $\beta_{1,2}$ and $\theta_{1,2}$ are defined in Eqs. \eqref{f7}, \eqref{f8}. Next, we aim to express $\Phi$ in terms of theta functions. From Eqs. \eqref{f10}, \eqref{f2} and \eqref{f5}, we could verify
    $(-2\sqrt{2}\ii\lambda \beta_1)^2=-2(\beta_1-v_1)(\beta_2-v_2)(\beta_3-v_3).$
    Moreover, from Eqs. \eqref{f10}, \eqref{f7}, \eqref{f8} and \eqref{f23}, we get
\begin{equation}\label{f34}
\begin{aligned}
     4\lambda^3(z)+2s_2\lambda+s_3\pm 4\lambda y =2y\left(\frac{\dd y}{\dd \lambda}\pm 2\lambda\right)=\frac{\sqrt{2}\alpha}{2K}\frac{\dd}{\dd z}\left(y(z)\pm \lambda^2(z)\right),
\end{aligned}
\end{equation}
which implies $2\lambda\beta_1=\sqrt{2}\alpha\beta_{1,z}/(4K)$ and then $\left( -\ii\alpha\beta_{1,z}/(2K)\right)^2=-2(\beta_1-v_1)(\beta_2-v_2)(\beta_3-v_3)$ holds. According to the existence and uniqueness theorem for the ordinary differential equation, we get
$\beta_1=2\alpha^2 k^2\left(\sn^2(4\ii l K)-\sn^2(2\ii(z-C)K) \right)$, 
where $C$ is an undetermined constant. In order to determine $C$, we plug $z=0$ into $\beta_1=2\lambda^2+s_2+2y$. From Eqs. \eqref{f23} and \eqref{lambda}, we obtain $C=l$ and hence we get
$
    \beta_1=2\lambda^2(z)+s_2+2y(z)=2\alpha^2 k^2\left(\sn^2(4\ii l K)-\sn^2(z_l)\right)$.
In the same way, we can also obtain
$
     \beta_2=2\lambda^2(z)+s_2-2y(z)=2\alpha^2 k^2\left(\sn^2(4\ii l K)-\sn^2(\ii K^{\prime}-z_l-4\ii lK)\right)$,
and thus we can obtain
\begin{equation}\label{f35}
\begin{aligned}
     &v(\xi)+\beta_1=2\alpha^2 k^2\left(\sn^2(\alpha \xi)-\sn^2(z_l)\right), \quad
     v(\xi)+\beta_2=2\alpha^2 k^2\left(\sn^2(\alpha \xi)-\sn^2(\ii K^{\prime}-4\ii l K-z_l)\right).
\end{aligned}
\end{equation}
In the following, we aim to transform them into theta functions. At this step, we utilize the same method in \cite{ling2023stability}, which avoids the tedious calculation. It is well known that the Jacobi elliptic functions are double periodic meromorphic functions. Therefore, the functions shown in Eq. \eqref{f35} have the period $2K/\alpha$ and $\ii K^{\prime}/\alpha$ with respect to $\xi$. So we only have to consider functions in the period area $\xi\in \left[-K/\alpha,K/\alpha \right]\times \left[0,\ii K^{\prime}/\alpha\right]$. We notice that $v+\beta_1$ could be rewritten as
$v+\beta_1=2\alpha^2k^2\left(\sn(\alpha \xi)+\sn(z_l) \right)\left(\sn(\alpha \xi)-\sn(z_l) \right)$.
From which, we could derive that $\xi=\pm z_l/\alpha$ are zeros of $v+\beta_1$ and $\xi=\ii K^{\prime}/\alpha$ is the double pole of $v+\beta_1$ based on the property of  the Jacobi elliptic functions. According to the Liouville Theorem, we have $ \left(\sn^2(\alpha \xi)-\sn^2(z_l) \right)=C_1\vartheta_1\left(\hat{\alpha}\xi +\ii(z-l) \right)\vartheta_1\left(\hat{\alpha}\xi -\ii(z-l) \right)/\vartheta_4^2\left(\ii(z-l)K\right)$, where $C_1$ is undermined. By setting $\xi=0$ in above equation, we get $C_1=\vartheta_3^2\vartheta_4^2(\hat{\alpha}\xi )/\vartheta_4^2(\ii(z-l))$ and thus the expression of $v+\beta_1$ in theta functions could be obtained. Similarly, we can derive the expression of $v+\beta_2$ in theta functions.
Thus $v+\beta_1$, $v+\beta_2$ have the following representation:
    \begin{equation}\label{f14}
        \begin{split}
            v+\beta_1&=2\alpha^2\frac{\vartheta_2^2\vartheta_4^2\vartheta_1(\hat{\alpha}\xi +\ii(z-l) )\vartheta_1(\hat{\alpha}\xi -\ii(z-l) )}{\vartheta_3^2\vartheta_4^2(\hat{\alpha}\xi )\vartheta_4^2(\ii(z-l) )},\quad
            v+\beta_2=-2\alpha^2\frac{\vartheta_2^2\vartheta_4^2\vartheta_4(\hat{\alpha}\xi +\ii(z-l) )\vartheta_4(\hat{\alpha}\xi -\ii(z-l) )}{\vartheta_3^2\vartheta_4^2(\hat{\alpha}\xi )\vartheta_1^2(\ii(z-l) )}.
        \end{split}
    \end{equation}

In order to express $\Phi_{11,12}$ in terms of theta functions, we first consider the integral terms in $\theta_{1,2}$ defined in Eq. \eqref{f8}. From Eq. \eqref{f34}, we have
$
     2\lambda\beta_1+s_3=-2\sqrt{2}\ii \alpha^3 k^2 \mathrm{scd}(z_l),\quad
    2\lambda\beta_2+s_3=-2\sqrt{2}\ii \alpha^3 k^2 \mathrm{scd}(\ii K^{\prime}-z_l-4\ii lK)$.
Through calculations analogous to those in the proof of Eq. \eqref{solu2}, we can obtain
\begin{equation}\label{f45}
    \begin{split}
        \frac{\sqrt{2}\ii}{2}\int_0^{\xi}\frac{2\lambda\beta_1+s_3}{v+\beta_1}\dd z&=\frac{1}{2}\ln \frac{\vartheta_1(\ii(z-l) -\hat{\alpha}\xi )}{\vartheta_1(\ii(z-l) +\hat{\alpha}\xi )}+\alpha Z(z_l)\xi,\\
        \frac{\sqrt{2}\ii}{2}\int_0^{\xi}\frac{2\lambda\beta_2+s_3}{v+\beta_2}\dd z&=\frac{1}{2}\ln \frac{\vartheta_1(\tau /2-\ii(z+l) -\hat{\alpha}\xi )}{\vartheta_1(\tau /2-\ii(z+l) +\hat{\alpha}\xi )}+\alpha Z(\ii K^{\prime}-z_l-4\ii l K)\xi.
    \end{split}
\end{equation}
On account of Eqs. \eqref{f34}, \eqref{f45}, denote the $(i,j)$-element of $\Phi$ as $\Phi_{ij}$, we can represent functions $\Phi_{11,12}$ in terms of the theta functions as follow:
\begin{equation}\label{f15}
    \Phi_{11}=\sqrt{2}\ii\alpha\frac{\vartheta_2\vartheta_4}{\vartheta_3}\frac{\vartheta_1\left(\hat{\alpha}\xi -\ii(z-l) \right)}{\vartheta_4(\ii(z-l) )\vartheta_4\left(\hat{\alpha}\xi \right)}E_1,\quad  \Phi_{12}=\sqrt{2}\ii\alpha\frac{\vartheta_2\vartheta_4}{\vartheta_3}\frac{\vartheta_4\left(\hat{\alpha}\xi +\ii(z+l) \right)}{\vartheta_1(\ii(z+l) )\vartheta_4\left(\hat{\alpha}\xi \right)}E_2,
\end{equation}
where $E_{1,2}=E_{1,2}(\xi,\eta;z)$ are defined in Eq. \eqref{psi}. Furthermore, in order to represent $\Phi_{21,22}$ in terms of the theta functions, we first rewrite $r_{1,2}$ in terms of the theta functions. According to the similar calculation in Eq. \eqref{f14}, from Eqs. \eqref{f23} and \eqref{lambda} we can obtain 
\begin{equation}\nonumber
    \lambda-\mu=\frac{\sqrt{2}\ii\alpha}{2\dn(4\ii l K)}\frac{\vartheta_4^2\vartheta_2\vartheta_1\left(\hat{\alpha}\xi -\ii(z-l) \right)\vartheta_3(-2\ii l  )\vartheta_4\left(-\hat{\alpha}\xi -\ii(z+l) \right)}{\vartheta_3^2\vartheta_1\left(-\hat{\alpha}\xi -2\ii l \right)\vartheta_1(-\ii(z+l) )\vartheta_4(\ii(z-l) )\vartheta_4\left(\hat{\alpha}\xi \right)},
\end{equation}

then we can express $(\lambda-\mu^{*})/(\lambda-\mu)$ as 
$$\frac{\lambda-\mu^{*}}{\lambda-\mu}=\frac{\vartheta_1\left(\hat{\alpha}\xi +\ii(z-l) \right)\vartheta_1\left(\hat{\alpha}\xi +2\ii l \right)\vartheta_4\left(\hat{\alpha}\xi -\ii(z+l) \right)}{\vartheta_1\left(\hat{\alpha}\xi -\ii(z-l) \right)\vartheta_1\left(\hat{\alpha}\xi -2\ii l \right)\vartheta_4\left(\hat{\alpha}\xi +\ii(z+l) \right)}.$$
As for $(v+\beta_2)/(v+\beta_1)$, from Eq. \eqref{f34}, it is easy to verify $(v+\beta_2)/(v+\beta_1)=\vartheta_4^2(\ii(z-l) )\vartheta_4\left(\hat{\alpha}\xi +\ii(z+l) \right)\vartheta_4\left(\hat{\alpha}\xi -\ii(z+l) \right)/(\vartheta_1^2(\ii(z+l) )\vartheta_1\left(\hat{\alpha}\xi +\ii(z-l) \right)\vartheta_1\left(\hat{\alpha}\xi -\ii(z-l) \right))$.
Meanwhile, from the presentation of $\psi(\xi,\eta)$ in Eq. \eqref{solu2} we have $\psi^*/\psi=\vartheta_1\left(-\hat{\alpha}\xi +2\ii l  \right)/\vartheta_1\left(\hat{\alpha}\xi +2\ii l  \right)\ee^{-2\omega_1\xi-2\omega_2\eta}$.
Moreover, by the definition of $r_{i}$, we have
\begin{equation}\nonumber
    r_1=\sqrt{\frac{y-f}{y+f}\cdot\frac{h}{g}}=\ii\sqrt{\frac{v+\beta_2}{v+\beta_1}\cdot\frac{\lambda-\mu^*}{\lambda-\mu}\cdot\frac{\psi^*}{\psi}},\quad  
    r_2=\sqrt{\frac{y+f}{y-f}\cdot\frac{h}{g}}=\ii\sqrt{\frac{v+\beta_1}{v+\beta_2}\cdot\frac{\lambda-\mu^*}{\lambda-\mu}\cdot\frac{\psi^*}{\psi}}.
\end{equation}
Combining the above equation, we can obtain
\begin{equation*}
    \begin{aligned}
        r_1&=\ii\frac{\vartheta_4(\ii(z-l) )\vartheta_{4}\left(\hat{\alpha}\xi -\ii(z+l) \right)}{\vartheta_{1}(-\ii(z+l) )\vartheta_{1}\left(\hat{\alpha}\xi -\ii(z-l) \right)}\ee^{-\omega_1\xi-\omega_2\eta}, \quad
        r_2=\ii\frac{\vartheta_1(\ii(z+l) )\vartheta_{1}\left(\hat{\alpha}\xi +\ii(z-l) \right)}{\vartheta_{4}(\ii(z-l) )\vartheta_{4}\left(\hat{\alpha}\xi +\ii(z+l) \right)}\ee^{-\omega_1\xi-\omega_2\eta}.
    \end{aligned}
\end{equation*}
Together with the above equation and Eq. \eqref{f15}, we can represent $\Phi_{21,22}(\xi,\eta;z)$ in terms of the theta functions,
\begin{equation}\label{f25}
\begin{split}
    \Phi_{21}&=-\sqrt{2}\alpha\frac{\vartheta_2\vartheta_4}{\vartheta_3}\frac{\vartheta_4\left(\hat{\alpha}\xi -\ii(z+l)\right)}{\vartheta_1(-\ii(z+l) )\vartheta_4\left(\hat{\alpha}\xi \right)}E_1\ee^{-\omega_1\xi-\omega_2\eta},\quad
    \Phi_{22}=-\sqrt{2}\alpha\frac{\vartheta_2\vartheta_4}{\vartheta_3}\frac{\vartheta_1\left(\hat{\alpha}\xi +\ii(z-l) \right)}{\vartheta_4(\ii(z-l) )\vartheta_4\left(\hat{\alpha}\xi \right)}E_2\ee^{-\omega_1\xi-\omega_2\eta}.
    \end{split}
\end{equation}
Combining Eqs. \eqref{f15} and \eqref{f25}, we can obtain Eq. \eqref{psi}.

\section{\appendixname. Proofs of Eq. \eqref{inf1}, \eqref{inf2} and \eqref{eq:psi-solution}}\label{app3}
\textbf{The proof of the Eqs. \eqref{inf1} and \eqref{inf2}:}
    In order to calculate the infimum, we start from $|\psi^{[1]}(\xi,\eta)|^2$. It is easy to derive that
    \begin{equation}
        |\psi^{[1]}(\xi,\eta)|^2=\frac{2\alpha^2\vartheta_2^2\vartheta_4^2}{\vartheta_3^2|\vartheta_4(2\ii l )|^2}\cdot \frac{|H|^2}{\left(\vartheta_4(\hat{\alpha}\xi +2z_c )\hat{E}_1+2\hat{a}_1\vartheta_4(\hat{\alpha}\xi )\vartheta_1(2z_c )\right)^2},
    \end{equation}
    where $|H|^2=|\vartheta_1(\hat{\alpha}\xi +2z_c +2\ii l )|^2\hat{E}_1^2+4\hat{a}_1\vartheta_1(2z_c )\hat{E}_1\Re\left(-p^{*}(z_1)p^{-1}(z_1)\vartheta_1(\hat{\alpha}\xi +2z_c +2\ii l  )\vartheta_1(\hat{\alpha}\xi -2\ii l  )\right)+4\hat{a}_1^2|\vartheta_1(\hat{\alpha}\xi +2\ii l )|^2\vartheta_1^2(2z_c )$, $\hat{E}_1$ is defined in Eq. \eqref{psi1}.
    Since $\psi^{[1]}(\xi,\eta)$ is an elliptic dark soliton solution, its background is periodically oscillating, which is determined by the theta functions. Therefore, when we analyze its infimum, we only need to analyze the infimum of the theta functions. 
    In view of the fact that the theta functions are bounded periodic functions, we have $|\vartheta_1(\hat{\alpha}\xi +2\ii l )|\in[|\vartheta_1(2\ii l )|,|\vartheta_2(2\ii l )|]$ and $\vartheta_4(\hat{\alpha}\xi +2z_c )\in[\vartheta_4,\vartheta_3]$ by utilizing the properties of theta functions (\cite{armitage2006elliptic} p.104). Moreover, $|\vartheta_1(\hat{\alpha}\xi +2z_c+2\ii l)|$ and $\vartheta_4(\hat{\alpha}\xi +2z_c)$ have the same supremum and infimum as $|\vartheta_1(\hat{\alpha}\xi+2\ii l)|$ and $\vartheta_4(\hat{\alpha}\xi)$ respectively. Meanwhile, there exists $d_m\in\mathbb{R}$ such that $d_m=\underset{{\xi\in\mathbb{R}}}{\mathrm{min}}(\Re\left(-p^{*}(z_1)p^{-1}(z_1)\vartheta_1(\hat{\alpha}\xi +2z_c +2\ii l  )\vartheta_1(\hat{\alpha}\xi -2\ii l  )\right))$.
    
    When $z_1\in S$, $2\hat{a}_1\vartheta_1(2z_c )>0$ holds, thus we could denote $2\rho_1=\ln(2\hat{a}_1\vartheta_1(2z_c ))$. To calculate its infimum, one can make the denominator take on the maximum value and the numerator take on the minimum value. Subsequently, it can be derived that
    \begin{equation}\nonumber
        \begin{split}
            |\psi^{[1]}(\xi,\eta)|^2&\geq 
            |\chi|^2\frac{|\vartheta_1(2\ii l )|^2\hat{E}_1^2+4\hat{a}_1^2|\vartheta_1(2\ii l )|^2\vartheta_1^2(2z_c )+4\hat{a}_1d_m\vartheta_1(2z_c )\hat{E}_1}{\vartheta_3^2\left(\hat{E}_1+2\hat{a}_1\vartheta_1(2z_c )\right)^2}\\
            &=
             |\chi|^2\frac{|\vartheta_1(2\ii l )|^2\left(\ee^{\hat{L}-\rho_1}+\ee^{-\hat{L}+\rho_1}\right)^2-2|\vartheta_1(2\ii l )|^2+2d_m}{\vartheta_3^2\left(\ee^{\hat{L}-\rho_1}+\ee^{-\hat{L}+\rho_1}\right)^2}\\
            &=
             \frac{|\chi\vartheta_1(2\ii l )|^2}{\vartheta_3^2}\left(1+\frac{2d_m-2|\vartheta_1(2\ii l )|^2}{|\vartheta_1(2\ii l )|^2}\cdot\frac{1}{\left(\ee^{\hat{L}-\rho_1}+\ee^{-\hat{L}+\rho_1}\right)^2}\right),
        \end{split}
    \end{equation}    
    where $\hat{E}_1$ and $\hat{L}$ are defined in Eqs. \eqref{psi1} and \eqref{inf2}.
    
    It can be noted that $d_m\leq |\vartheta_1(2\ii l )|^2$ when $z_1 \in S$ with $\Re(z_1)=\ii\tau/4$. Through calculation, we can obtain a constant $B_1=(d_{m}+\ii \sqrt{|\vartheta_1(2\ii l )|^4-d_m^2})/|\vartheta_1(2\ii l )|^2$ such that
    \begin{equation}\nonumber
        \begin{split}
            |\psi^{[1]}(\xi,\eta)|^2\geq 
            \frac{|\chi\vartheta_1(2\ii l )|^2}{\vartheta_3^2}\left(\frac{B_1-1}{2}+\frac{B_1+1}{2}\tanh(\hat{L}-\rho_1)\right)\left(\frac{B_1^{*}-1}{2}+\frac{B_1^{*}+1}{2}\tanh(\hat{L}-\rho_1)\right).
        \end{split}
    \end{equation}
    
    On the other hand, when $z_1\in S$ with $\Re(z_1)=-\ii\tau/4$, there exists a constant $\tilde{d}_m\geq |\vartheta_1(2\ii l )|^2$ satisfying 
    \begin{equation}
        |\psi^{[1]}(\xi,\eta)|^2
        \geq
         \frac{|\chi\vartheta_1(2\ii l )|^2}{\vartheta_3^2}\left(1+\frac{2\tilde{d}_m-2|\vartheta_1(2\ii l )|^2}{|\vartheta_1(2\ii l )|^2}\cdot\frac{1}{\left(\ee^{\hat{L}-\rho_1}+\ee^{-\hat{L}+\rho_1}\right)^2}\right).
    \end{equation}
     Since $d_m$ is estimated to be as small as possible, we might not obtain a greatest infimum. In order to make the infimum closer to the greatest infimum, we set $\tilde{d}_{m}=\mathrm{max}[d_m,|\vartheta_1(2\ii l )|^2]$.  In this case, there exists a pure imaginary constant $B_2=\ii\sqrt{(\tilde{d}_{m}-|\vartheta_1(2\ii l )|^2)/(2|\vartheta_1(2\ii l )|^2})$ such that
     \begin{equation}\nonumber
        \begin{split}
            |\psi^{[1]}(\xi,\eta)|^2\geq 
             \frac{|\chi\vartheta_1(2\ii l )|^2}{\vartheta_3^2}\left(1+B_2\sech(\hat{L}-\rho_1)\right)\left(1+B_2^{*}\sech(\hat{L}-\rho_1)\right).
        \end{split}
    \end{equation}
Above all, the infimum of $\psi^{[1]}(\xi,\eta)$ with $\Re(z_1)=\pm\ii\tau/4$ could be obtained as Eqs. \eqref{inf1} and \eqref{inf2} respectively. The proof of the supremum is similar to that of the infimum. It can be achieved by making the denominator take the minimum value and the numerator take the maximum value.

\textbf{The proof of the Eq. \eqref{eq:psi-solution}:}
Similar to the proof of Eq. \eqref{f14}, by analyzing the poles and zeros, we have
\begin{equation}\label{f11}
    \lambda_j-\lambda_i^{*}=\lambda(z_j)-\lambda(z_i^{*})=
    \frac{\chi\vartheta_4^2(2il )\vartheta_1(\ii(z_i^{*}-z_j) )\vartheta_4(\ii(z_i^{*}+z_j) )}{2\vartheta_1(\ii(z_i^{*}+l) )\vartheta_1(\ii(z_j+l) )\vartheta_4(\ii(z_j-l) )\vartheta_4(\ii(z_i^{*}-l) )}.
\end{equation}
From Eq. \eqref{psi}, we can present $\Phi_i^{\dagger}$ and $\Phi_j$ as
\begin{align*}
    \Phi_i^{\dagger}&=
    \frac{\chi^* \vartheta_4\left(2\ii l \right)}{\vartheta_4\left(\hat{\alpha}\xi \right)}\mathbf{c}_i^{\dagger}\mathbf{E}_i^{\dagger} \begin{bmatrix}
        -\frac{\vartheta_1\left(-\hat{\alpha}\xi -\ii(z_i^{*}-l) \right)}{\vartheta_4(-\ii(z_i^{*}-l) )}&
        -\frac{\vartheta_4\left(-\hat{\alpha}\xi -\ii(z_i^{*}+l) \right)}{\vartheta_1(-\ii(z_i^{*}+l) )}\\
        -\frac{\vartheta_4\left(-\hat{\alpha}\xi +\ii(z_i^{*}+l) \right)}{\vartheta_1(\ii(z_i^{*}+l) )}&
        -\frac{\vartheta_1\left(-\hat{\alpha}\xi +\ii(z_i^{*}-l) \right)}{\vartheta_4(\ii(z_i^{*}-l) )}
    \end{bmatrix}\Lambda^{\dagger},\quad
    \Phi_j=
    \frac{\chi \vartheta_4\left(2\ii l \right)}{\vartheta_4\left(\hat{\alpha}\xi \right)}\Lambda\begin{bmatrix}
        \frac{\vartheta_1\left(\hat{\alpha}\xi -\ii(z_j-l) \right)}{\vartheta_4(-\ii(z_j-l) )}&
        \frac{\vartheta_4\left(\hat{\alpha}\xi +\ii(z_j+l) \right)}{\vartheta_1(\ii(z_j+l) )}\\
        \frac{\vartheta_4\left(\hat{\alpha}\xi -\ii(z_j+l) \right)}{\vartheta_1(-\ii(z_j+l) )}&
        \frac{\vartheta_1\left(\hat{\alpha}\xi +\ii(z_j-l) \right)}{\vartheta_4(\ii(z_j-l) )}
    \end{bmatrix}\mathbf{E}_j\mathbf{c}_j.
\end{align*}
Since $\Lambda^{\dagger}\sigma_3\Lambda=\sigma_3$, we can express $\Phi_i^{\dagger}\sigma_3\Phi_j$ as follows by using the addition formula (\cite{KHARCHEV201519} p.25),
\begin{equation}\label{f12}
\Phi_i^{\dagger}\sigma_3\Phi_j=\frac{2\alpha^2\vartheta_2^2\vartheta_4^2\vartheta_4(2\ii l )}{\vartheta_3^2\vartheta_4\left(\hat{\alpha}\xi \right)}\triangle^{-1}\mathbf{c}_i^{\dagger}\mathbf{E}_i^{\dagger}\begin{bmatrix}
       A_{11}&A_{12}\\
        A_{21}&-A_{11}
    \end{bmatrix}\mathbf{E}_j\mathbf{c}_j,
\end{equation}
with $A_{11,12,21}$ and $\triangle$ defined as:
\begin{equation*}
    \begin{aligned}  
     A_{11}&=\vartheta_4\left(\hat{\alpha}\xi +\ii(z_i^{*}-z_j) \right)\vartheta_4(\ii(z_i^{*}+z_j) ),\quad
  A_{12}=\vartheta_1\left(\hat{\alpha}\xi +\ii(z_i^{*}+z_j) \right)\vartheta_1(\ii(z_i^{*}-z_j) ),\\
    A_{21}&=\vartheta_1\left(\hat{\alpha}\xi -\ii(z_i^{*}+z_j) \right)\vartheta_1(\ii(z_i^{*}-z_j) ),\quad
  \triangle=\vartheta_1(\ii(z_i^{*}+l) )\vartheta_1(\ii(z_j+l) )\vartheta_4(\ii(z_i^{*}-l) )\vartheta_4(\ii(z_j-l) ).
    \end{aligned}
\end{equation*}
By combining Eq. \eqref{f11} and Eq. \eqref{f12}, we have
\begin{equation}\label{f30}
    \frac{\Phi_i^{\dagger}\sigma_3\Phi_j}{2(\lambda_j-\lambda_i^{*})}
    =
    -\chi\frac{\vartheta_4\left(2\ii l \right)}{\vartheta_4\left(\hat{\alpha}\xi \right)}\mathbf{c}_i^{\dagger} \mathbf{E}_i^{\dagger}\begin{bmatrix}
        \frac{\vartheta_4\left(\hat{\alpha}\xi +\ii(z_i^*-z_j) \right)}{\vartheta_1(\ii(z_i^*-z_j) )}&
        \frac{\vartheta_1\left(\hat{\alpha}\xi +\ii(z_i^*+z_j) \right)}{\vartheta_4(\ii(z_i^*+z_j) )}\\
        \frac{\vartheta_1\left(\hat{\alpha}\xi -\ii(z_i^*+z_j) \right)}{\vartheta_4(-\ii(z_i^*+z_j) )}&
        \frac{\vartheta_4\left(\hat{\alpha}\xi -\ii(z_i^*-z_j) \right)}{\vartheta_1(-\ii(z_i^*-z_j) )}\end{bmatrix}\mathbf{E}_j\mathbf{c}_j.
\end{equation}
Furthermore, we derive the expression of $\psi\mathbf{M}_N-\ii \mathbf{X}_{N,2}^{\dagger}\mathbf{X}_{N,1}$ by using the the addition formula \cite{KHARCHEV201519}. We can obtain
\begin{equation}\label{f31}
    \begin{split}
       & \psi\frac{\Phi_i^{\dagger}\sigma_3 \Phi_j}{2(\lambda_j-\lambda_i^{*})}-\ii \mathbf{X}_{N,2}^{\dagger}\mathbf{X}_{N,1}\\
        =&\frac{2\alpha^2 \vartheta_2^2\vartheta_4^2}{\vartheta_3^2\vartheta^2_4\left(\hat{\alpha}\xi \right)}\mathbf{c}_i^{\dagger}\mathbf{E}_i^{\dagger}\left(\frac{\vartheta_1(\hat{\alpha}\xi +2\ii l )}{\vartheta_4(2\ii l  )}\begin{bmatrix}
        \frac{\vartheta_4\left(\hat{\alpha}\xi +\ii(z_i^*-z_j) \right)}{\vartheta_1(\ii(z_i^*-z_j) )}&
        \frac{\vartheta_1\left(\hat{\alpha}\xi +\ii(z_i^*+z_j) \right)}{\vartheta_4(\ii(z_i^*+z_j) )}\\
        \frac{\vartheta_1\left(\hat{\alpha}\xi -\ii(z_i^*+z_j) \right)}{\vartheta_4(-\ii(z_i^*+z_j) )}&
        \frac{\vartheta_4\left(\hat{\alpha}\xi -\ii(z_i^*-z_j) \right)}{\vartheta_1(-\ii(z_i^*-z_j) )}\end{bmatrix}\right.\\
        &+\left.\begin{bmatrix}
        \frac{\vartheta_4\left(\hat{\alpha}\xi +\ii(z_i^*+l) \right)\vartheta_1\left(\hat{\alpha}\xi -\ii(z_j-l) \right)}{\vartheta_1(-\ii(z_i^*+l) )\vartheta_4(\ii(z_j-l) )}&
        \frac{\vartheta_4\left(\hat{\alpha}\xi +\ii(z_i^*+l) \right)\vartheta_4\left(\hat{\alpha}\xi +\ii(z_j+l) \right)}{\vartheta_1(-\ii(z_i^*+l) )\vartheta_1(\ii(z_j+l) )}\\
       \frac{\vartheta_1\left(-\hat{\alpha}\xi +\ii(z_i^*+l) \right)\vartheta_1\left(\hat{\alpha}\xi -\ii(z_j-l) \right)}{\vartheta_4(\ii(z_i^*-l) )\vartheta_4(\ii(z_j-l) )}&
        \frac{\vartheta_1\left(-\hat{\alpha}\xi +\ii(z_i^*-l) \right)\vartheta_4\left(\hat{\alpha}\xi +\ii(z_j+l) \right)}{\vartheta_4(\ii(z_i^*-l) )\vartheta_1(\ii(z_j-l) )}\end{bmatrix}\right)
       \mathbf{E}_j\mathbf{c}_j\ee^{\omega_1\xi+\omega_2\eta}\\
        =&\frac{-2\alpha^2 \vartheta_2^2\vartheta_4^2}{\vartheta_3^2\vartheta_4\left(\hat{\alpha}\xi \right)\vartheta_4(2\ii l  )}\mathbf{c}_i^{\dagger}\mathbf{E}_i^{\dagger}\mathbf{p}_i^{\dagger}
    \begin{bmatrix}
        \frac{\vartheta_1\left(\hat{\alpha}\xi +\ii(z_i^*-z_j) +2\ii l  \right)}{\vartheta_1(\ii(z_i^*-z_j) )}&
        \frac{\vartheta_4\left(\hat{\alpha}\xi +\ii(z_i^*+z_j) +2\ii l  \right)}{\vartheta_4(\ii(z_i^*+z_j) )}\\
        \frac{\vartheta_4\left(\hat{\alpha}\xi -\ii(z_i^*+z_j) +2\ii l  \right)}{\vartheta_4(-\ii(z_i^*+z_j) )}&
        \frac{\vartheta_1\left(\hat{\alpha}\xi -\ii(z_i^*-z_j) +2\ii l  \right)}{\vartheta_1(-\ii(z_i^*-z_j) )}\end{bmatrix}\mathbf{p}_j^{-1}\mathbf{E}_j\mathbf{c}_j\ee^{\omega_1\xi+\omega_2\eta},
        \end{split}
\end{equation}
where $ \mathbf{p}_i=\mathbf{p}(z_i)=\diag(p(z_i),-p^{-1}(z_i))$ with $ p(z_i)=\vartheta_4(\ii(z_i-l))/\vartheta_1(-\ii(z_i+l))$.

  Since $z_i\in S$ defined in Eq. \eqref{f16}, we notice that when $\mathrm{Re}(z_i)=\pm \ii \tau/4$, there has $\vartheta_4(\ii(z_i+z_i^{*}) )=0$. Based on Eq. \eqref{f11}, $\lambda_i-\lambda_i^{*}$ becomes zero when $\vartheta_4(\ii(z_i+z_i^{*}) )=0$. To address this issue, we choose such a parameter $c_{i2}$ that can eliminate the zero generated in $\lambda_i-\lambda_i^{*}$ and thus we set $c_{i2}=a_i \vartheta_4(\ii(z_i+z_i^{*}) )$. From Eq. \eqref{f30}, we have
\begin{equation}\label{f17}
\begin{aligned}
    \frac{\Phi_i^{\dagger}\sigma_3\Phi_j}{2(\lambda_j-\lambda_i^{*})}
    =&
    \frac{-\chi\vartheta_4\left(2\ii l \right)}{\vartheta_4\left(\hat{\alpha}\xi \right)}\begin{bmatrix}
        1&a_i^{*}
    \end{bmatrix}\mathbf{E}_i^{\dagger}\begin{bmatrix}
        \frac{\vartheta_4\left(\hat{\alpha}\xi +\ii(z_i^*-z_j) \right)}{\vartheta_1(\ii(z_i^*-z_j) )}&
        \delta_{ij}\vartheta_1\left(\hat{\alpha}\xi +\ii(z_i^*+z_j) \right)\\
        \delta_{ij}\vartheta_1\left(\hat{\alpha}\xi -\ii(z_i^*+z_j) \right)&
        0\end{bmatrix}\mathbf{E}_j
\begin{bmatrix}
        1\\
        a_j
    \end{bmatrix}, 
\end{aligned}
\end{equation}
where $\delta_{ij}$ is Kronecker delta.
Moreover, we have $E_1(z_i)E_2^{*}(z_i)=\mathrm{exp}(\ii\hat{\alpha}\xi )$, $E_1^{*}(z_i)E_2(z_i)=\mathrm{exp}(-\ii\hat{\alpha}\xi )$, then we can deduce $a_i^{*}\vartheta_1\left(\hat{\alpha}\xi -\ii(z_i^*+z_i) \right)E_1(z_i)E_2^{*}(z_i)+a_i\vartheta_1\left(\hat{\alpha}\xi +\ii(z_i^*+z_i) \right)E_1^{*}(z_i)E_2(z_i)
    =a_i^{*}\ii\ee^{-\ii\tau /4}\vartheta_4\left(\hat{\alpha}\xi \right)-a_i\ii\ee^{-\ii\tau /4}\vartheta_4\left(\hat{\alpha}\xi \right)
    =2\mathrm{Im}(a_i)\ee^{-\ii\tau /4}\vartheta_4\left(\hat{\alpha}\xi \right)$.
Combining it with Eq. \eqref{f17}, we can obtain a simpler form of $(\mathbf{M}_{N})_{ij}$ as:
\begin{equation}\label{f19}
    \frac{\Phi_i^{\dagger}\sigma_3\Phi_j}{2(\lambda_j-\lambda_i^{*})}=
    -2\hat{a}_i\chi\vartheta_4(2\ii l)\left(\frac{\vartheta_4\left(\hat{\alpha}\xi +\ii(z_i^{*}-z_j) \right)E_1^{*}(z_i)E_1(z_j)}{2\hat{a}_i\vartheta_1(\ii(z_i^{*}-z_j) )\vartheta_4\left(\hat{\alpha}\xi \right)}+\delta_{ij}\right),
\end{equation}
where $\hat{a}_i=\mathrm{Im}(a_i)\ee^{-\ii\tau /4}$. In a similar way, form Eq. \eqref{f31} we can obtain
\begin{equation*}
    \begin{split}
        &\psi\frac{\Phi_i^{\dagger}\sigma_3 \Phi_j}{2(\lambda_j-\lambda_i^{*})}-\ii \mathbf{X}_{N,2}^{\dagger}\mathbf{X}_{N,1}\\
    =&\frac{-2\alpha^2\vartheta_2^2\vartheta_4^2}{\vartheta_3^2\vartheta_4\left(\hat{\alpha}\xi \right)\vartheta_4(2\ii l  )}\begin{bmatrix}
        1&a_i^{*}
    \end{bmatrix}\mathbf{E}_i^{\dagger}\mathbf{p}_i^{\dagger}
    \begin{bmatrix}
        \frac{\vartheta_1\left(\hat{\alpha}\xi +\ii(z_i^*-z_j) +2\ii l  \right)}{\vartheta_1(\ii(z_i^*-z_j) )}&
        \delta_{ij}\vartheta_4\left(\hat{\alpha}\xi +\ii(z_i^*+z_j) +2\ii l  \right)\\
        \delta_{ij}\vartheta_4\left(\hat{\alpha}\xi -\ii(z_i^*+z_j) +2\ii l  \right)&
        0\end{bmatrix}\mathbf{p}_j^{-1}\mathbf{E}_j\begin{bmatrix}
        1\\
        a_j
    \end{bmatrix}\ee^{\omega_1\xi+\omega_2\eta}\\
 =&\frac{4\alpha^2\hat{a}_i\vartheta_2^2\vartheta_4^2\vartheta_1\left(\hat{\alpha}\xi +2\ii l \right)}{\vartheta_3^2\vartheta_4\left(\hat{\alpha}\xi \right)\vartheta_4(2\ii l  )}\left(\frac{p^{*}(z_i)\vartheta_1\left(\hat{\alpha}\xi +\ii(z_i^{*}-z_j) +2\ii l \right)E_1^{*}(z_i)E_1(z_j)}{-2\hat{a}_ip(z_j)\vartheta_1(\ii(z_i^{*}-z_j) )\vartheta_1\left(\hat{\alpha}\xi +2\ii l  \right)}+\delta_{ij}\right)\ee^{\omega_1\xi+\omega_2\eta},
    \end{split}
\end{equation*}
where we use $p^{*}(z_i)p(z_i)=\ee^{2l }$ in the last equality. Combining the above equations with Eq. \eqref{darb}, $\psi^{[N]}(\xi,\eta)$ could be derived. 

To ensure $\psi^{[N]}(\xi,\eta)$ is analytical for any $(\xi,\eta)\in \mathbb{R}^2$, we would like to prove $\mathrm{det}(\mathcal{E}^{\dagger}\mathcal{G}\mathcal{E}+\mathbb{I}_N)>0$ holds for all $(\xi,\eta)\in \mathbb{R}^2$. We denote the $N\times N$ sequential principal minors of the matrix $\mathcal{G}$ as $(\mathcal{G})_N$, and we have
\begin{equation*}
    \mathrm{det}((\mathcal{G})_N)=\frac{\vartheta_4\left(\hat{\alpha}\xi +\sum_{i=1}^{N}\ii(z_i^{*}-z_i) \right)\prod_{1\leq i<j\leq N}\vartheta_1(\ii(z_i^{*}-z_j^{*}) )\vartheta_1(-\ii(z_i-z_j) )}{2^N\vartheta_4(\hat{\alpha}\xi )\prod_{i=1}^{N}\hat{a_i}\prod_{i,j=1}^{N}\vartheta_1(\ii(z_i^{*}-z_j) )}.
\end{equation*}
Without loss of generality, we assume that the first $h$ number of parameters $z_i$, $i=1,2,\cdots,h$ satisfy $\mathrm{Re}(z_i)=-\ii\tau/4$, while the last number of parameters $z_i$, $i=h+1,h+2,\cdots,N$ satisfy $\mathrm{Re}(z_i)=\ii\tau/4$. From which, we could deduce that $\vartheta_1(\ii(z^{*}_i-z^{*}_j) )\vartheta_1(-\ii(z_i-z_j) )=\vartheta^2_1(\mathrm{Im}(z_i-z_j) )$ when $1\leq i<j\leq h$ or $h+1\leq i< j\leq N$. Furthermore, we could obtain $ \prod_{1\leq i\leq h<j\leq N}\vartheta_1(\ii(z^{*}_i-z^{*}_j) )\vartheta_1(-\ii(z_i-z_j) )=\prod_{1\leq i\leq h<j\leq N}\vartheta_1(\mathrm{Im}(z_i-z_j) +\tau /2)\vartheta_1(\mathrm{Im}(z_i-z_j) -\tau /2)=\prod_{1\leq i\leq h<j\leq N}\vartheta^2_4(\mathrm{Im}(z_i-z_j) )\ee^{-\ii\tau /2}$.
On the other hand, in a similar way, we have $ \prod^N_{i,j=1}\vartheta_1(\ii(z^{*}_i-z_j) )=\prod^N_{i=1}\vartheta_1(\ii(z^{*}_i-z_i) )\prod_{1\leq i\leq h<j\leq N}\vartheta_1(\ii(z^{*}_i-z_j) )\vartheta_1(\ii(z^{*}_j-z_i) )=\prod^N_{i=1}\vartheta_1(2\mathrm{Im}(z_i) )\prod_{1\leq i <j\leq N}\vartheta_4(\mathrm{Im}(z_i+z_j) )\ee^{-\ii\tau /2}$.
Based on above equations, we could simplify $\mathrm{det}((\mathcal{G})_N)$ as follows,
\begin{equation*}
    \begin{split}
        \mathrm{det}((\mathcal{G})_N)=\frac{\vartheta_4\left(\hat{\alpha}\xi +\sum_{i=1}^{N}\ii(z_i^{*}-z_i) \right)\prod_{1\leq i<j\leq N}\vartheta^2_4(\mathrm{Im}(z_i-z_j) )}{2^N\vartheta_4(\hat{\alpha}\xi )\prod_{i=1}^{N}\hat{a_i}\vartheta_1(2\mathrm{Im}(z_i) )\prod_{1\leq i<j\leq N}\vartheta^2_4(\mathrm{Im}(z_i+z_j) )}.
    \end{split}
\end{equation*}
In order to obtain $\mathrm{det}((\mathcal{G})_N)>0$, we need to prove that the inequality $\prod_{i=1}^{N}\hat{a_i}\vartheta_1(2\mathrm{Im}(z_i) )>0$ holds. When $z_i\in S$, $2\mathrm{Im}(z_i)\in (0,1)$, which implies $\vartheta_1(2\mathrm{Im}(z_i) )>0$. Therefore, since $\hat{a}_i>0$, we have  $\prod_{i=1}^{N}\hat{a_i}\vartheta_1(2\mathrm{Im}(z_i) )>0$. Above all, we could conclude that $\psi^{[N]}(\xi,\eta)$ is analytic for all $(\xi,\eta)\in \mathbb{R}^2$ with the fact that $\mathrm{det}(\mathcal{E}^{\dagger}\mathcal{G}\mathcal{E}+\mathbb{I}_N)>0$.

\textbf{The proof of the Eq. \eqref{psi1}:}
The proof is analogous to that of Eq. \eqref{eq:psi-solution}. One can refer to the proof of Eq. \eqref{eq:psi-solution} by setting $N=1$, and we will not elaborate on it here.
By taking $N=1$, we can obtain $\psi^{[1]}(\xi,\eta)$ from Eq. \eqref{eq:psi-solution}. On one hand, we have
    \begin{equation}\nonumber        \mathcal{E}^{\dagger}\mathcal{G}\mathcal{E}+\mathbb{I}_1=\frac{\vartheta_4(\hat{\alpha}\xi+\ii(z_1^{*}-z_1))}{2\hat{a}_1\vartheta_1(\ii(z_1^{*}-z_1))\vartheta_4(\hat{\alpha}\xi)}|E_1(z_1)|^2+1.
    \end{equation}
    On the other hand,
    \begin{equation}\nonumber    \mathcal{E}^{\dagger}\mathcal{P}^{\dagger}\mathcal{H}\mathcal{P}^{-1}\mathcal{E}+\mathbb{I}_1=\frac{p^{*}(z_1)\vartheta_1(\hat{\alpha}\xi+\ii(z_1^{*}-z_1)+2\ii l)}{-2\hat{a}_1p(z_1)\vartheta_1(\ii(z_1^{*}-z_1))\vartheta_1(\hat{\alpha}\xi+2\ii l)}|E_1(z_1)|^2+1.
    \end{equation}
    Based on the definition of $E_1(z_1)$ given by Eq. \eqref{psi}, we have $|E_1(z_1)|^2=\exp(\alpha (Z(z_l)+Z(z_l^*))\xi+2\ii(\sqrt{2}\lambda-1)y\eta)$. Applying the addition formula for the Jacobi Zeta function (\cite{takahashi2016integrable}, p.34), we can obtain $Z(z_l)+Z(z_l^*)=-Z(4\Im(z_1)K)-k^2\sn(\pm\ii K^{\prime}/2+2\ii l K-2\Im(z_1)K)\sn(\pm\ii K^{\prime}/2+2\ii l K+2\Im(z_1)K)\sn(2\Im(z_1)K)$ with $\Re(z_1)=\pm\ii\tau/4$.
    Combining the above equation with Eq. \eqref{eq:psi-solution}, the explicit expression of $\psi^{[1]}$ could be derived.

\textbf{The proof of the Eq. \eqref{f50}:}
         From the definition of $D_i$, we have $D_j=\Re(\tilde{W}_j)D_i/\Re(\tilde{W}_i)+\Re(\tilde{W}_j)(\nu_i-\nu_j) \eta$, $j\neq i$. As $\eta\rightarrow +\infty$, along the line $D_i$, we get $D_j\rightarrow +\infty$, $j<i$ and $D_j\rightarrow-\infty$, $j>i$ since $\nu_1<\nu_2<\cdots <\nu_N$.
	The solution provided in Eq. \eqref{eq:psi-solution} could be rewritten as
	\begin{equation}\nonumber
        \begin{split}
			\psi^{[N]}(\xi,\eta)=&
            \frac{\chi\vartheta_1(\hat{\alpha} \xi +2\ii l )}{\vartheta_4(\hat{\alpha} \xi )}\cdot\frac{\det(\mathcal{F}^{\dagger}\mathcal{E}^{\dagger}\mathcal{P}^{\dagger}\mathcal{H}\mathcal{P}^{-1}\mathcal{E}\mathcal{F}+\mathcal{F}^{\dagger}\mathbb{I}_N\mathcal{F})}{\det(\mathcal{F}^{\dagger}\mathcal{E}^{\dagger}\mathcal{G}\mathcal{E}\mathcal{F}+\mathcal{F}^{\dagger}\mathbb{I}_N\mathcal{F})}\ee^{\omega_1\xi+\omega_2\eta},
            \end{split}
		\end{equation}
		where $\mathcal{F}=\diag\left(\ee^{-W_1\xi-V_1\eta},
		\cdots, \ee^{-W_{i-1}\xi-V_{i-1}\eta}, 1, \cdots,1\right)$.
		For $\eta\rightarrow +\infty$, along the trajectories $D_i$, the asymptotic expression of the solution $\psi^{[N]}_+(\xi,\eta;D_i)$ is expressed as 
		\begin{equation}\nonumber
			\begin{split}
				\psi^{[N]}_+(\xi,\eta;D_i)
				\rightarrow&
				 \frac{\chi\vartheta_1(\hat{\alpha} \xi +2\ii l )}{\vartheta_4(\hat{\alpha} \xi )}\cdot\frac{\det(\mathbf{h}_{i})\ee^{2D_i}+\det(\mathbf{h}_{i-1})}
				{\det(\mathbf{g}_{i})\ee^{2D_i}+\det(\mathbf{g}_{i-1})}\ee^{\omega_1\xi+\omega_2\eta},
			\end{split}
		\end{equation}
		where matrices $\mathbf{h}_{i}$ and $\mathbf{g}_{i}$ are defined as 
        \begin{equation}\nonumber
			\begin{split}
				\mathbf{h}_{i}
				=&\left(\frac{-\vartheta_1(\hat{\alpha}\xi +\ii(z_m^*-z_n) +2\ii l )p^*_m}{2\hat{a}_m\vartheta_1(\ii(z_m^*-z_n) )\vartheta_1(\hat{\alpha}\xi +2\ii l )p_m}
				\right)_{1\le m,n\le i},\qquad
				\mathbf{g}_{i}
				=\left(\frac{\vartheta_4(\hat{\alpha}\xi +\ii(z_m^*-z_n) )}{2\hat{a}_m\vartheta_1(\ii(z_m^*-z_n) )\vartheta_4(\hat{\alpha}\xi )}\right)_{1\le m,n\le i}.
			\end{split}
		\end{equation}
		Combined with the determinant of theta functions \cite{takahashi2016integrable},
		it is easy to obtain that 
        \begin{equation}
			\begin{split}
				\det(\mathbf{g}_i)=&\frac{\vartheta_4(\hat{\alpha}\xi +\sum_{j=1}^{i}\ii(z_j^*-z_j) )\prod_{1\le m<n\le i}\vartheta_1(\ii (z_m^*-z_n^*) )\vartheta_1(-\ii (z_m-z_n) )}{2^i\vartheta_4\left(\hat{\alpha}\xi \right)\prod_{m=1}^{i}\hat{a}_m\prod_{m,n=1}^{i}\vartheta_1(\ii (z_m^*-z_n) )},\\
				\det(\mathbf{h}_i)=&\mathbf{r}^{+}_{i+1}\frac{\vartheta_1(\hat{\alpha}\xi +\sum_{j=1}^{i}\ii(z_j^*-z_j) +2\ii l )\prod_{1\le m<n\le i}\vartheta_1(\ii (z_m^*-z_n^*) )\vartheta_1(-\ii (z_m-z_n) )}{(-2)^i\vartheta_1\left(\hat{\alpha}\xi +2\ii l \right)\prod_{m=1}^{i}\hat{a}_m\prod_{m,n=1}^{i}\vartheta_1(\ii (z_m^*-z_n) )}.
			\end{split}
		\end{equation}
		Then, $\psi^{[N]}_{+}(\xi,\eta;D_i)$ could be simplified as
        \begin{equation}\nonumber
	\psi^{[N]}_{+}(\xi,\eta;D_i)\rightarrow(-1)^i\frac{\sqrt{2}\ii\alpha\vartheta_2\vartheta_4}{\vartheta_3\vartheta_4(2\ii l )}\mathbf{r}^{+}_i\frac{\frac{\vartheta_1(\hat{\alpha}\xi +\mathbf{s}_i^{+}+\ii(z_i^*-z_i) +2\ii l )}{2\hat{a}_i\vartheta_1(\ii (z_i^*-z_i) )}\ee^{2D_i+\gamma_i^{+}}-\vartheta_1(\hat{\alpha}\xi +\mathbf{s}_i^{+}+2\ii l )}	{\frac{\vartheta_4(\hat{\alpha}\xi +\mathbf{s}_i^{\pm}+\ii(z_i^*-z_i) )}{2\hat{a}_i\vartheta_1(\ii (z_i^*-z_i) )}\ee^{2D_i}+\vartheta_4(\hat{\alpha}\xi +\mathbf{s}_i^{+})}\ee^{\omega_1\xi+\omega_2\eta},
\end{equation}
		where parameters $\mathbf{s}_i^+$, $\mathbf{r}_i^+$, and $\gamma_i^+$ are defined in Eq. \eqref{eq:s-r-gamma-define}.
		Similarly, as $\eta \rightarrow -\infty$, we could obtain the exact expression of $\psi^{[N]}_{-}(\xi,\eta;D_i)$. In summary, the exact expression of the solution $\psi^{[N]}_{\pm}(\xi,\eta;D_i)$ holds.

\end{widetext}

\bibliography{your-bib-file}

\end{document}